\documentclass[aps,prd,twocolumn,amsmath,showpacs,superscriptaddress,nofootinbib,nopreprintnumbers,tightenlines,notitlepage]{revtex4-1}

\usepackage{verbatim}
\usepackage[T1]{fontenc}
\usepackage[utf8]{inputenc}
\usepackage[american]{babel}
\usepackage{epsfig}
\usepackage{graphicx}
\usepackage{hyperref}
\usepackage{cleveref}
\usepackage[justification=centerlast]{caption}

\usepackage{booktabs}
\usepackage{multirow}
\usepackage{dcolumn}
\usepackage{amsmath}
\usepackage{mathtools}
\usepackage{amsfonts}
\usepackage{amssymb}
\usepackage{epstopdf}
\usepackage{bm}
\usepackage{siunitx}
\usepackage{braket}
\usepackage{enumitem}
\usepackage{soul}
\usepackage[table]{xcolor}
\usepackage{color}
\usepackage{transparent}

\usepackage{enumitem}
\usepackage{graphicx}
\usepackage[font=small]{caption}
\usepackage{subcaption}
\usepackage{pifont}
\usepackage[normalem]{ulem}

\definecolor{navyblue}{rgb}{0.0, 0.0, 0.5}
\definecolor{royalblue}{rgb}{0.25, 0.41, 0.88}
\definecolor{cadmiumgreen}{rgb}{0.0, 0.42, 0.24}
\definecolor{blue-violet}{rgb}{0.54, 0.17, 0.89}
\definecolor{darkviolet}{rgb}{0.58, 0.0, 0.83}
\definecolor{teal(colorwheel)}{rgb}{1.0, 0.5, 0.0}

\usepackage{hyperref}
\hypersetup{
    colorlinks=true, % false: boxed links; true: colored links
    linkcolor=royalblue, % color of internal links (change box color with linkbordercolor)
    citecolor=magenta}

\newcommand\ee{\end{equation}}
\newcommand\be{\begin{equation}}
\newcommand\eea{\end{eqnarray}}
\newcommand\bea{\begin{eqnarray}}

\usepackage{booktabs}
\usepackage{multirow}
\usepackage{dcolumn}
\usepackage{colortbl}

\definecolor{magenta(process)}{rgb}{1.0, 0.0, 0.56}

\definecolor{darkspringgreen}{rgb}{0.09, 0.45, 0.27}

\definecolor{royalblue(web)}{rgb}{0.25, 0.41, 0.88}

\begin{document}

\title{Extended Analysis of Neutrino-Dark Matter Interactions with Small-Scale CMB Experiments}

\author{Philippe Brax}
\email{philippe.brax@ipht.fr}
\affiliation{Institut de Physique Th\'eorique, Universit\'e  Paris-Saclay, CEA, CNRS, F-91191 Gif-sur-Yvette Cedex, France}
\affiliation{CERN, Theoretical Physics Department, Geneva, Switzerland.}

\author{Carsten van de Bruck}
\email{c.vandebruck@sheffield.ac.uk}
\affiliation{School of Mathematics and Statistics, University of Sheffield, Hounsfield Road, Sheffield S3 7RH, United Kingdom}

\author{Eleonora Di Valentino}
\email{e.divalentino@sheffield.ac.uk}
\affiliation{School of Mathematics and Statistics, University of Sheffield, Hounsfield Road, Sheffield S3 7RH, United Kingdom}

\author{William Giar\`e}
\email{w.giare@sheffield.ac.uk}
\affiliation{School of Mathematics and Statistics, University of Sheffield, Hounsfield Road, Sheffield S3 7RH, United Kingdom}

\author{Sebastian Trojanowski}
\email{Sebastian.Trojanowski@ncbj.gov.pl}
\affiliation{Astrocent, Nicolaus Copernicus Astronomical Center Polish Academy of Sciences, ul.~Rektorska 4, 00-614, Warsaw, Poland}
\affiliation{National Centre for Nuclear Research, ul.~Pasteura 7, 02-093 Warsaw, Poland}

%\date{\today}

%\preprint{}

%%%%%%%%%%%%%%%%%%%%%%%%%%%%%%%%%%%%%%%%%%%%%%%%%%%%%%%%%%%%%%%%%%%%%
\begin{abstract}

We explore an extension of the standard $\Lambda$CDM model by including an interaction between neutrinos and dark matter, and making use of the ground based telescope data of the Cosmic Microwave Background (CMB) from the Atacama Cosmology Telescope (ACT). An indication for a non-zero coupling between dark matter and neutrinos (both assuming a temperature independent and $T^2$ dependent cross-section) is obtained at the 1$\sigma$ level coming from the ACT CMB data alone and when combined with the Planck CMB and Baryon Acoustic Oscillations (BAO) measurements. This result is confirmed  by both fixing the effective number of relativistic degrees of freedom in the early Universe to the Standard Model value of $N_{\rm eff}=3.044$, and allowing $N_{\rm eff}$ to be a free cosmological parameter. Furthermore, when performing a Bayesian model comparison, the interacting $\nu$DM (+$N_{\rm eff}$) scenario is mostly preferred over a baseline $\Lambda$CDM (+$N_{\rm eff}$) cosmology. 
The preferred value is then used as a benchmark and  the potential implications of dark matter's interaction with a sterile neutrino are discussed.

\end{abstract}

\maketitle

\section{Introduction}
A vast number of astrophysical and cosmological observations point towards the existence of dark matter (DM). DM dominates the matter content of the universe and yet can only interact very weakly with normal matter, if at all. The nature of DM and its properties remain unknown, but it is widely believed that DM consists of non-relativistic (‘cold’) massive particles. The strongest constraints on properties of DM come from cosmological observations and direct searches~\cite{Roszkowski:2017nbc,Billard:2021uyg,Buen-Abad:2021mvc}. 

One intriguing possibility which has been widely discussed in the literature is an interaction between DM and neutrinos~\cite{Bertoni:2014mva,GonzalezMacias:2015rxl,Gonzalez-Macias:2016vxy,Escudero:2016tzx,Escudero:2016ksa,Batell:2017cmf,Blennow:2019fhy}. Constraints on $\nu$DM couplings can be obtained by a variety astrophysical, cosmological or accelerator-based experiments~\cite{Boehm:2000gq,Boehm:2004th,Mangano:2006mp,Palomares-Ruiz:2007trf,Serra:2009uu,Shoemaker:2013tda,Wilkinson:2013kia,Bringmann:2013vra,Wilkinson:2014ksa,Farzan:2014gza,Escudero:2015yka,Shoemaker:2015qul,deSalas:2016svi,Arguelles:2017atb,Batell:2017rol,DiValentino:2017oaw,Olivares-DelCampo:2017feq,Escudero:2018thh,Pandey:2018wvh,Kelly:2018tyg,Kelly:2019wow,Alvey:2019jzx,Choi:2019ixb,Jho:2021rmn,Ghosh:2021vkt,Kelly:2021mcd,Lin:2022dbl,Cline:2022qld,Ferrer:2022kei,Cline:2023tkp}. Cosmological tests of such interactions include studies of CMB anisotropies, probes of the LSS power spectrum and Lyman-$\alpha$ data. Theories with $\nu$DM interactions predict a suppression of perturbations at length scales smaller than the collisionless damping scale. As a consequence, structure formation on smaller length scales differ from the uncoupled case. It has been pointed out that $\nu$DM interactions potentially address shortcomings of standard CDM models,  such as the missing satellites or the too-big-to-fail problems~\cite{Shoemaker:2013tda,Boehm:2014vja,Bertoni:2014mva,Schewtschenko:2015rno}. In the presence of a $\nu$DM coupling, DM perturbations entering the horizon in the radiation dominated epoch no longer simply grow via gravitational instability. Instead, the coupled DM--neutrino fluid experiences damped oscillations, similar to the coupled photon-baryon fluid. Neutrinos no longer free-stream because they are bound to the dynamics of the DM particles. One consequence of these processes is a change of the CMB anisotropy power spectrum. We refer to \cite{Serra:2009uu,Wilkinson:2014ksa} for detailed discussions on the physical processes involved. 

We have recently pointed out that measurements of CMB anisotropies on small angular scales (large multipole numbers $\ell$) with a few percent accuracy provide a significant amount of information on $\nu$DM interactions, while their imprint on larger angular scales, as those probed e.g.~by Planck, would require much larger sensitivity~\cite{Brax:2023rrf}. In this work we will continue our study of $\nu$DM interactions on the CMB anisotropies at large multipoles ($l \gtrsim 3000$), emphasising the role of recent and future CMB data at high multipoles to constrain such interactions. We will consider mainly the case of a temperature independent cross-section, but briefly comment on the case of a $T^2$ dependent cross-section as well. Following the literature, we quantify the interaction by the parameter
\begin{equation}
	u_{\nu \textrm{DM}} = \frac{\sigma_{\nu\textrm{DM}}}{\sigma_{\textrm{T}}}\,\left(\frac{m_{\textrm{DM}}}{100~\textrm{GeV}}\right)^{-1},
	\label{eq:defu}
\end{equation}
where $\sigma_{\textrm{T}}$ is the Thomson scattering cross section and $m_{\rm DM}$ is the mass of the DM particle. 

\begin{figure}
    \centering
    \includegraphics[width=\columnwidth]{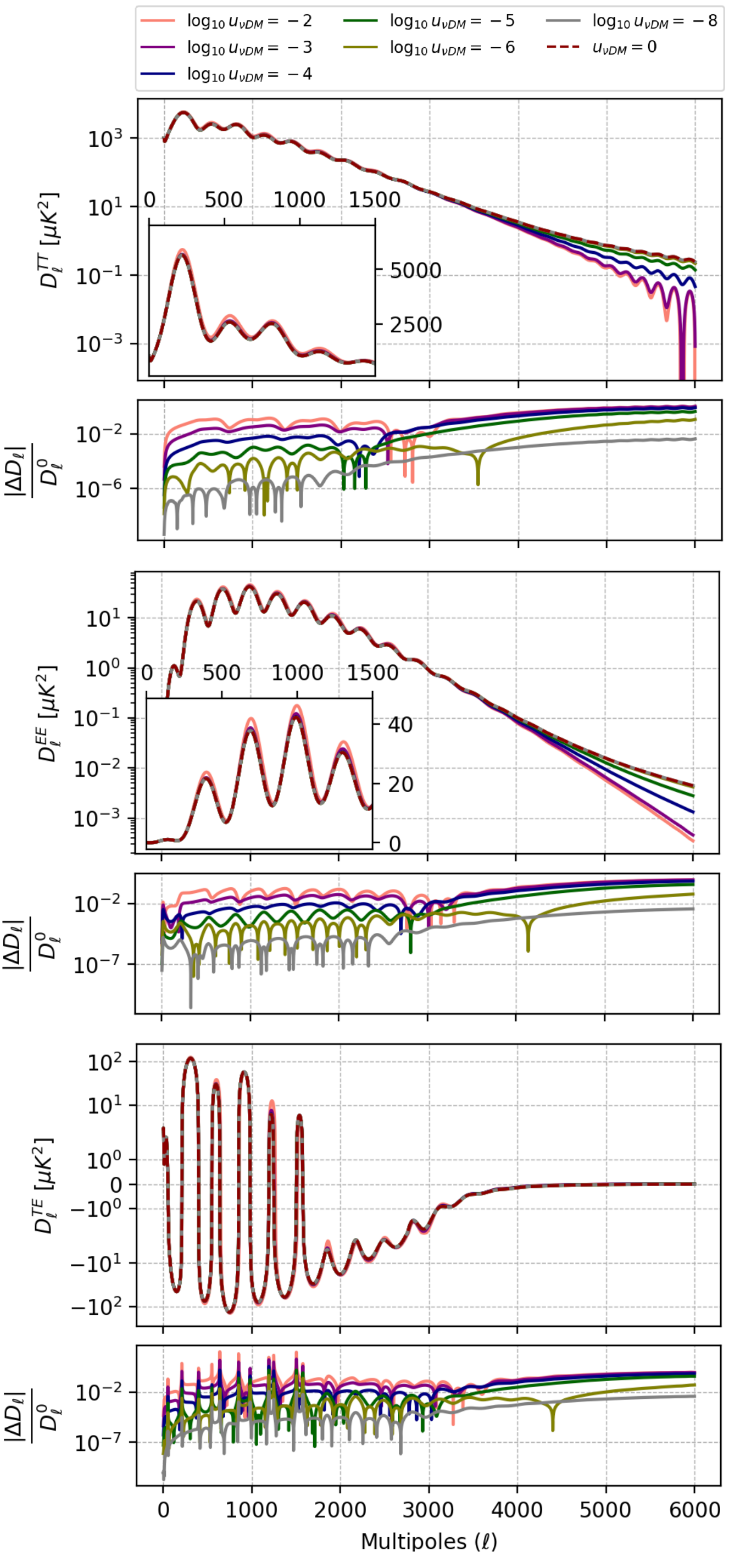}
    \caption{Theoretical temperature (TT) and polarization (EE and TE) angular power spectra $D_{\ell}^{XX}$ (top panel) and the percentage difference $|\Delta D_{\ell}| / D_{\ell}^{0}$ with respect to the non interacting case ($D_{\ell}^{0}$) for different values of the coupling.}
    \label{fig:theory}
\end{figure}

The paper is organized as follows. In~\autoref{sec:method} we describe the codes and data used for the analysis, in~\autoref{sec:results} we present our constrains for the cases explored in this paper, in~\autoref{sec:modelexample} we show the impact of our results on a specific model of $\nu$DM interaction via a sterile neutrino portal, and finally in~\autoref{sec:concl} we draw our conclusions.

\section{Methods} \label{sec:method}

\begin{table}
	\begin{center}
		\renewcommand{\arraystretch}{1.5}
		\begin{tabular}{c@{\hspace{0. cm}}@{\hspace{0.7 cm}} c @{\hspace{0.7 cm}} c }
			\hline
			\textbf{Parameter}                       & $\sigma_{\nu\rm{DM}}\sim T^0$  & $\sigma_{\nu\rm{DM}}\sim T^2$\\
			\hline\hline
			$\Omega_{\rm b} h^2$                     & $[0.005\,,\,0.1]$ &$[0.005\,,\,0.1]$\\
			$\Omega_{\rm c}^{\nu\rm DM} h^2$         & $[0.005\,,\,0.1]$ &$[0.005\,,\,0.1]$\\
			$100\,\theta_{\rm {MC}}$                 & $[0.5\,,\,10]$ &$[0.5\,,\,10]$\\
			$\tau$                                   & $[0.01\,,\,0.8]$ & $[0.01\,,\,0.8]$\\
			$\log(10^{10}A_{\rm S})$                 & $[1.61\,,\,3.91]$ & $[1.61\,,\,3.91]$\\
			$n_{s}$                                  & $[0.8\,,\, 1.2]$ & $[0.8\,,\, 1.2]$\\
                $N_{\rm eff}$                            & $[0\,,\,10]$ & $[0\,,\,10]$\\
			$\log_{10}u_{\nu \rm{DM}}$                  & $[-8\,,\,-1]$ & $[-18\,,\,-12]$\\
			
			\hline\hline
		\end{tabular}
		\caption{List of the uniform parameter priors.}
		\label{tab.Priors}
	\end{center}
	
\end{table}

In this paper we examine extended cosmological models that include interactions between DM and neutrinos to determine the constraints that can be obtained from the latest Cosmic Microwave Background (CMB) and large-scale structure probes. To this end, we make use of the publicly available code \texttt{COBAYA}~\cite{Torrado:2020xyz}. The code explores the posterior distributions of a given parameter space using the Monte Carlo Markov Chain (MCMC) sampler developed for \texttt{CosmoMC}~\cite{Lewis:2002ah} and tailored for parameter spaces with speed hierarchy implementing the “fast dragging” procedure detailed in Ref.~\cite{Neal:2005}. To compute the theoretical model and introduce the possibility of interactions between neutrinos and DM, we exploit a modified versions of the Cosmic Linear Anisotropy Solving System code, \texttt{CLASS}~\cite{Blas:2011rf}.\footnote{A  publicly available version of this modified \texttt{CLASS} can be found at \url{https://github.com/MarkMos/CLASS_nu-DM} , see also Refs.~\cite{Stadler:2019dii,Mosbech:2020ahp}.} We treat neutrinos as massless and ultra-relativistic in the early universe. This approximation is widely used in literature and simplifies calculations for scenarios involving interactions with DM. In our analysis, we take into account the interaction between neutrinos and the entire fraction of DM energy density. Our baseline sampling considers the six usual $\Lambda$CDM parameters, namely the baryon $\omega_{\rm b}\doteq \Omega_{\rm b}h^2$ and cold dark matter $\omega_{\rm c}^{\nu\text{DM}}\doteq\Omega_{\rm c}^{\nu\text{DM}}h^2$ energy densities, the angular size of the horizon at the last scattering surface $\theta_{\rm{MC}}$, the optical depth $\tau$, the amplitude of primordial scalar perturbation $\log(10^{10}A_{\rm s})$ and the scalar spectral index $n_s$.  In addition, we consider the logarithm of the coupling parameter $\log_{10}u_{\nu \rm{DM}}$ defined in \autoref{eq:defu}, exploring two scenarios: a temperature-independent $\nu$DM cross-section ($\sigma_{\nu\rm{DM}}\sim T^0$) and a squared-temperature-dependent cross-section ($\sigma_{\nu\rm{DM}}\sim T^2$). In both cases, we begin by setting the effective number of ultra-relativistic particles at recombination ($N_{\rm eff}$) to its reference value of $N_{\rm eff}=3.044$~\cite{Akita:2020szl,Froustey:2020mcq,Bennett:2020zkv}. We then allow $N_{\rm eff}$ to be an additional free parameter of the sample, enabling us to relax this requirement and explore a wider range of possibilities. The prior distributions for all the sampled parameters involved in our analysis are chosen to be uniform along the range of variation provided in \autoref{tab.Priors}, with the only exception of the optical depth at reionization ($\tau$) for which the prior distribution is chosen accordingly to the CMB datasets. The convergence of the chains obtained with this procedure is tested using the Gelman-Rubin criterion~\cite{Gelman:1992zz} and we choose a threshold for chain convergence of $R-1 \lesssim 0.02 $.   

\begin{table*}
\begin{center}
\renewcommand{\arraystretch}{1.5}
\resizebox{\textwidth}{!}{
\begin{tabular}{l c c c c c c c c c c c c c c c }
\hline
\textbf{Parameter} & \textbf{ Planck } & \textbf{ Planck + BAO } & \textbf{ ACT } & \textbf{ ACT + BAO } & \textbf{ ACT + Planck + BAO } \\ 
\hline\hline
$ \Omega_\mathrm{b} h^2  $ & $  0.02239\pm 0.00015 $ & $  0.02239\pm 0.00013 $ & $  0.02153\pm 0.00030 $ & $  0.02154\pm 0.00030 $ & $  0.02236\pm 0.00012 $ \\ 
$ \Omega_\mathrm{c}^{\nu\mathrm{DM}} h^2  $ & $  0.1196\pm 0.0012 $ & $  0.11958\pm 0.00093 $ & $  0.1185\pm 0.0039 $ & $  0.1198\pm 0.0015 $ & $  0.11975\pm 0.00097 $ \\ 
$ 100\theta_\mathrm{s}  $ & $  1.04193\pm 0.00030 $ & $  1.04191\pm 0.00028 $ & $  1.04337\pm 0.00069 $ & $  1.04321\pm 0.00063 $ & $  1.04206\pm 0.00026 $ \\ 
$ \tau_\mathrm{reio}  $ & $  0.0528\pm 0.0074 $ & $  0.0524\pm 0.0072 $ & $  0.064\pm 0.015 $ & $  0.062\pm 0.014 $ & $  0.0563\pm 0.0064 $ \\ 
$ \log(10^{10} A_\mathrm{s})  $ & $  3.039\pm 0.014 $ & $  3.038\pm 0.014 $ & $  3.049\pm 0.030 $ & $  3.047\pm 0.030 $ & $  3.053\pm 0.013 $ \\ 
$ n_\mathrm{s}  $ & $  0.9642\pm 0.0044 $ & $  0.9642\pm 0.0038 $ & $  1.004\pm 0.016 $ & $  1.001\pm 0.014 $ & $  0.9678\pm 0.0036 $ \\ 
$ log_{10}u_{\nu DM}  $ & $ < -4.42\, (< -3.95 ) $ & $ < -4.46\, (< -4.39 ) $ & $  -5.08^{+1.5}_{-0.98}\, (< -3.74 ) $ & $  -4.86^{+1.5}_{-0.83}\, (< -3.70 ) $ & $  -5.20^{+1.2}_{-0.74}\, (< -4.17 ) $ \\ 
$ H_0  $ & $  68.03\pm 0.55\, ( 68.0^{+1.1}_{-1.1} ) $ & $  68.05\pm 0.42\, ( 68.05^{+0.81}_{-0.82} ) $ & $  68.2\pm 1.6\, ( 68.2^{+3.3}_{-3.3} ) $ & $  67.66\pm 0.58\, ( 67.7^{+1.1}_{-1.2} ) $ & $  68.01\pm 0.43\, ( 68.01^{+0.83}_{-0.85} ) $ \\ 
$ \sigma_8  $ & $  0.806^{+0.013}_{-0.0097}\, ( 0.806^{+0.024}_{-0.028} ) $ & $  0.807^{+0.011}_{-0.0084}\, ( 0.807^{+0.020}_{-0.021} ) $ & $  0.823^{+0.025}_{-0.021}\, ( 0.823^{+0.046}_{-0.050} ) $ & $  0.821^{+0.025}_{-0.020}\, ( 0.821^{+0.044}_{-0.050} ) $ & $  0.820^{+0.011}_{-0.0093}\, ( 0.820^{+0.021}_{-0.023} ) $ \\ 
\hline
%$\Delta \chi^2$ & $ -10.4 $ & $ -10.3 $ & $ -3.38 $ & $ -3.63 $ & $ -6.43 $ \\
$\ln BF$ & $ -3.74 $ & $ -2.48 $ & $ -0.194 $ & $ -0.156 $ & $ 0.525 $ \\

\hline \hline
\end{tabular} }
\end{center}
\caption{\textbf{Temperature independent cross section}: We report the 68\% (95\%) CL constraints/bounds on the cosmological parameters above the line, while below the line we have the improvement of the $\chi^2$ of the best fit and the Bayes Factor, with respect to the $\Lambda$CDM scenario. }
\label{tab1}
\end{table*}

\begin{table*}
\begin{center}
\renewcommand{\arraystretch}{1.5}
\resizebox{\textwidth}{!}{
\begin{tabular}{l c c c c c c c c c c c c c c c }
\hline
\textbf{Parameter} & \textbf{ Planck } & \textbf{ Planck + BAO } & \textbf{ ACT } & \textbf{ ACT + BAO } & \textbf{ ACT + Planck + BAO } \\ 
\hline\hline
$ \Omega_\mathrm{b} h^2  $ & $  0.02230\pm 0.00022 $ & $  0.02233\pm 0.00018 $ & $  0.02102\pm 0.00045 $ & $  0.02116\pm 0.00040 $ & $  0.02210\pm 0.00017 $ \\ 
$ \Omega_\mathrm{c}^{\nu\mathrm{DM}} h^2  $ & $  0.1180\pm 0.0030 $ & $  0.1181\pm 0.0028 $ & $  0.1101\pm 0.0065 $ & $  0.1104\pm 0.0059 $ & $  0.1151\pm 0.0025 $ \\ 
$ 100\theta_\mathrm{s}  $ & $  1.04217\pm 0.00052 $ & $  1.04214\pm 0.00048 $ & $  1.0448\pm 0.0012 $ & $  1.0445\pm 0.0011 $ & $  1.04279\pm 0.00046 $ \\ 
$ \tau_\mathrm{reio}  $ & $  0.0521\pm 0.0076 $ & $  0.0522\pm 0.0069 $ & $  0.060\pm 0.015 $ & $  0.062\pm 0.014 $ & $  0.0548\pm 0.0066 $ \\ 
$ \log(10^{10} A_\mathrm{s})  $ & $  3.033\pm 0.017 $ & $  3.034\pm 0.015 $ & $  3.021\pm 0.036 $ & $  3.027\pm 0.033 $ & $  3.036\pm 0.016 $ \\ 
$ n_\mathrm{s}  $ & $  0.9599\pm 0.0084 $ & $  0.9614\pm 0.0066 $ & $  0.956\pm 0.034 $ & $  0.972\pm 0.023 $ & $  0.9562\pm 0.0068 $ \\ 
$ N_\mathrm{eff}  $ & $  2.93\pm 0.19\, ( 2.93^{+0.37}_{-0.37} ) $ & $  2.96\pm 0.16\, ( 2.96^{+0.32}_{-0.32} ) $ & $  2.36\pm 0.43\, ( 2.36^{+0.87}_{-0.81} ) $ & $  2.52\pm 0.33\, ( 2.52^{+0.65}_{-0.63} ) $ & $  2.74\pm 0.15\, ( 2.74^{+0.30}_{-0.29} ) $ \\ 
$ log_{10}u_{\nu  \rm{DM}}  $ & $ < -4.47\, (< -4.01 ) $ & $ < -4.48\, (< -4.06 ) $ & $  -4.77^{+1.5}_{-0.93}\, (< -3.53 ) $ & $  -5.08^{+1.6}_{-0.98}\, (< -3.77 ) $ & $  -5.29^{+1.3}_{-0.81}\, (< -4.21 ) $ \\ 
$ H_0  $ & $  67.3\pm 1.4\, ( 67.3^{+2.8}_{-2.8} ) $ & $  67.5\pm 1.1\, ( 67.5^{+2.1}_{-2.1} ) $ & $  63.0\pm 3.6\, ( 63^{+7}_{-7} ) $ & $  64.8\pm 1.9\, ( 64.8^{+3.7}_{-3.7} ) $ & $  66.1\pm 1.0\, ( 66.1^{+2.0}_{-2.1} ) $ \\ 
$ \sigma_8  $ & $  0.803^{+0.013}_{-0.012}\, ( 0.803^{+0.025}_{-0.026} ) $ & $  0.804^{+0.012}_{-0.011}\, ( 0.804^{+0.022}_{-0.024} ) $ & $  0.791\pm 0.030\, ( 0.791^{+0.058}_{-0.064} ) $ & $  0.798\pm 0.023\, ( 0.798^{+0.046}_{-0.047} ) $ & $  0.807\pm 0.011\, ( 0.807^{+0.022}_{-0.023} ) $ \\ 
\hline
%$\Delta \chi^2$ & $ -9.72 $ & $ -10 $ & $ -3.74 $ & $ -3.48 $ & $ -6.35 $ \\
$\ln BF$ & $ -3.01 $ & $ -2.37 $ & $ -0.707 $ & $ -0.157 $ & $ -1.43 $ \\
\hline \hline
\end{tabular} }
\end{center}
\caption{\textbf{Temperature independent cross section with \boldmath{$N_{\rm eff}$}}: We report the 68\% (95\%) CL constraints/bounds on the cosmological parameters above the line, while below the line we have the improvement of the $\chi^2$ of the best fit and the Bayes Factor, with respect to the $\Lambda$CDM+$N_{\rm eff}$ scenario.}
\label{tab2}
\end{table*} 

Concerning CMB and large-scale structure probes, our baseline data sets consist of:

\begin{itemize}

\item The full \emph{Planck 2018} temperature and polarization likelihood~\cite{Planck:2019nip,Planck:2018vyg,Planck:2018nkj}, in combination with the Planck 2018 lensing likelihood~\citep{Planck:2018lbu}, constructed from measurements of the power spectrum of the lensing potential. We refer to this dataset as “Planck”. 

\item The full \emph{Atacama Cosmology Telescope} temperature and polarization DR4 likelihood~\citep{ACT:2020frw}, assuming a conservative Gaussian prior on $\tau=0.065\pm0.015$. We refer to this dataset as "ACT”.

\item  The full \emph{Atacama Cosmology Telescope} temperature and polarization DR4 likelihood~\citep{ACT:2020frw}, in combination with the \emph{Planck 2018 TT TE EE} likelihood~\cite{Planck:2019nip,Planck:2018vyg,Planck:2018nkj} in the multipole range $2\le \ell \le 650$. We refer to this dataset as "ACT+Planck”. 

\item The Baryon Acoustic Oscillations (BAO) and Redshift Space Distortions (RSD) measurements from BOSS DR12~\citep{BOSS:2012dmf}. We refer to this dataset as "BAO”.

\end{itemize}

\begin{figure}
    \centering
    \includegraphics[width=0.95\columnwidth]{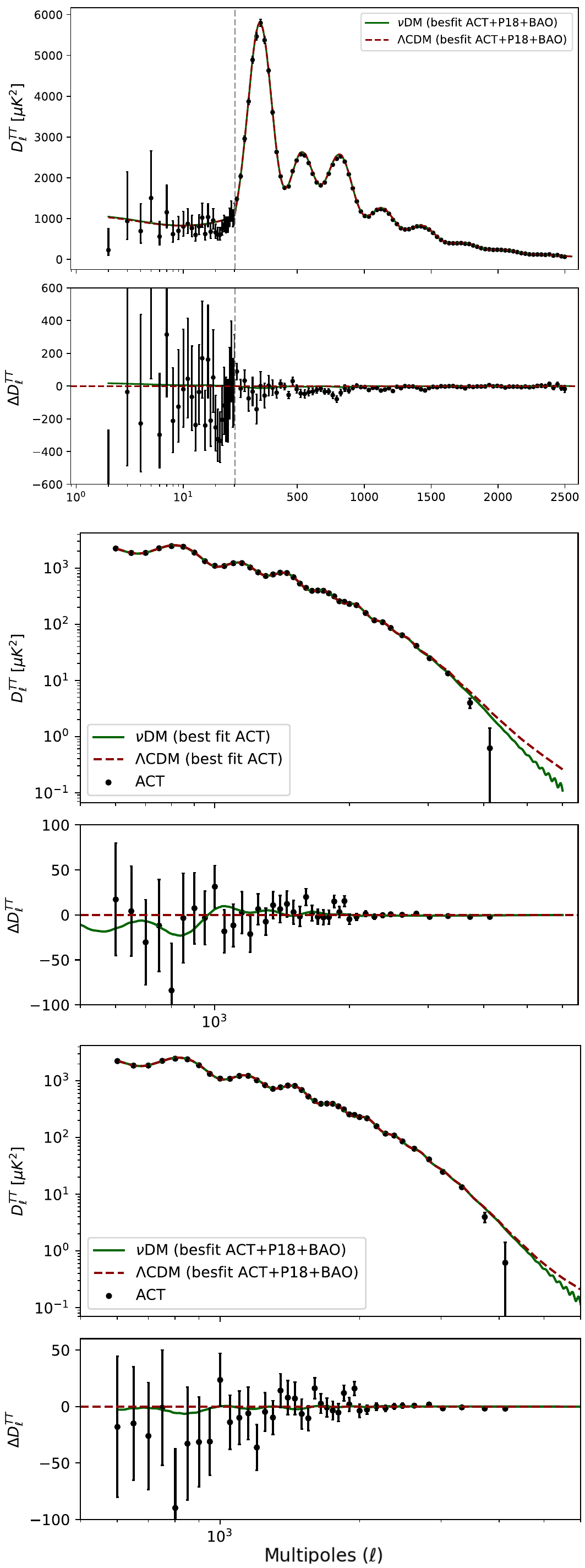}
    \caption{Comparison of the Planck, ACT and Planck plus ACT best fit temperature angular power spectra for $\nu$DM and $\Lambda$CDM cosmologies (with residuals in the lower panels). }
    \label{fig:residuals}
\end{figure}

Finally, to conduct a model comparison, we calculate the Bayesian evidence of each model and then estimate the corresponding Bayes factors $\ln \mathcal{B}{ij}$ using the publicly available package \texttt{MCEvidence}~\cite{Heavens:2017hkr,Heavens:2017afc},\footnote{The package is accessible at \url{https://github.com/yabebalFantaye/MCEvidence}.} that has been suitably modified to be compatible with \texttt{COBAYA}. In particular, for the case with $N_{\rm eff}$ fixed to $N_{\rm eff}=3.044$, we calculate $\ln \mathcal{B}{ij}$ as the difference between the evidence for $\Lambda$CDM and the interacting $\nu$DM cosmology ($\ln \mathcal{B}{ij} = \ln \mathcal{Z}_{\rm LCDM} - \ln \mathcal{Z}_{\rm \nu\rm DM}$). When instead $N_{\rm eff}$ is allowed to vary, we compute $\ln \mathcal{B}{ij}$ as the difference between the evidence for $\Lambda$CDM+$N{\rm eff}$ and the respective interacting scenario ($\ln \mathcal{B}{ij} = \ln \mathcal{Z}_{\rm LCDM+Neff} - \ln \mathcal{Z}_{\nu\rm{DM+Neff}}$). Within this convention, negative $\ln \mathcal{B}{ij}$ values indicate a preference for an interacting dark sector, while positive values indicate a preference for $\Lambda$CDM or $\Lambda$CDM+$N{\rm eff}$. We consider the evidence to be inconclusive if $0 \leq | \ln B_{ij}| < 1$, weak if $1 \leq | \ln B_{ij}| < 2.5$, moderate if $2.5 \leq | \ln B_{ij}| < 5$, strong if $5 \leq | \ln B_{ij}| < 10$, and very strong if $| \ln B_{ij} | \geq 10$, following the modified Jeffreys' scale~\cite{Jeffreys:1939xee,Trotta:2008qt}.

\section{Results} \label{sec:results}

In \autoref{tab1} we report the constraints on the cosmological parameters at 68\% (and 95\% CL) for the temperature independent $\nu$DM cross-section case when $N_{\rm eff}$ is fixed to its standard value $3.044$~\cite{Mangano:2005cc,Akita:2020szl,Froustey:2020mcq,Bennett:2020zkv}. We defer the discussion about the $T^2$ case to \hyperref[sec.T2]{Appendix A}.

The first thing to be  noticed is that there is simply an upper bound on the coupling between DM and neutrinos defined in \autoref{eq:defu}, using a logarithmic sampling, i.e. we are in complete agreement with a model without interaction. In particular we have that $\log_{10}u_{\nu \rm{DM}}<-4.42$ ($\log_{10}u_{\nu \rm{DM}}<-3.95$) at 68\% (95\%) CL for the Planck alone case, that is very consistent with the bounds we can find in the literature~\cite{DiValentino:2017oaw,Mosbech:2020ahp,Paul:2021ewd}, and robust after the inclusion of the BAO data. 
The interacting model is moderately favored by a Bayesian model comparison against the standard model with no interaction, as we can see from the value of the Bayes Factor reported in the last row of \autoref{tab1}.

However, it is when we analyze the alternative CMB data obtained by the ground based telescope ACT that we find the most important result of our paper, as already anticipated in Ref.~\cite{Brax:2023rrf}. Indeed we have an indication for a non-vanishing coupling between DM and neutrinos at slightly more than 68\% CL, namely $\log_{10}u_{\nu \rm{DM}}=-5.08^{+1.5}_{-0.98}$ at 1$\sigma$. Despite the fact that the indication vanishes at 95\% CL, we still consider this result significant because it does not arise from the typical discrepancies between various CMB experiments~\cite{Handley:2020hdp,DiValentino:2022rdg,DiValentino:2022oon, Calderon:2023obf}, i.e. producing controversial results on the extensions of the standard cosmological scenario,\footnote{See for example the results on Early Dark Energy~\cite{Hill:2021yec,Poulin:2023lkg}, inflationary parameters~\cite{Forconi:2021que,Giare:2022rvg}, curvature of the universe~\cite{DiValentino:2019qzk}, etc.} but rather to a better fit of the very small scales of the ACT data, that are beyond the range measured by Planck~\cite{Brax:2023rrf}. Actually, as can be seen in the middle panel of \autoref{fig:residuals} with the interaction between DM and neutrinos, the fit of the data points at high multipoles that are lower than the prediction of the $\Lambda$CDM scenario is improved as the coupling prevents neutrinos from free-streaming until they are decoupled from the DM and therefore increasing the damping of the fluctuations in that regime (see \autoref{fig:theory}). This 1$\sigma$ indication for a $\nu$DM coupling is very robust under the assumption of different priors for the parameter $\log_{10}u_{\nu \rm{DM}}$ sampled in our analysis, and the inclusion of additional large scale structure data such as BAO. 

\begin{figure}
    \centering
    \includegraphics[width=\columnwidth]{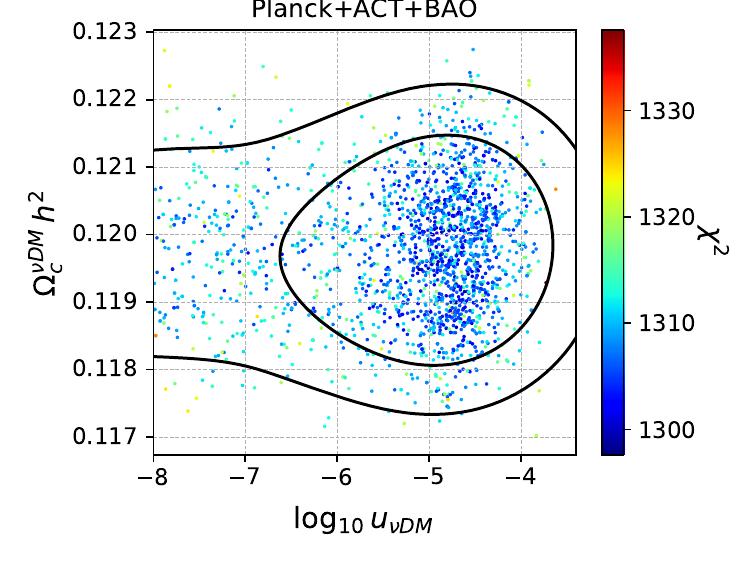}
    \caption{Distribution of $\chi^2$ values for different models of the samples in the two-dimensional plane ($\log_{10} u_{\nu \rm{DM}}$, $\Omega_c^{\nu DM}\, h^2$). The color gradient indicates the magnitude of the $\chi^2$ of each model's fit to the data, with red colors indicating higher values (i.e., worse fits). The fit to data generally improves around the peak of the posterior distribution of $\log_{10} u_{\nu \rm{DM}}$.}
    \label{fig:dist_chi2}
\end{figure}

In order to prove that this indication is not due to the absence of the first peak in the ACT data and the failure in breaking the possible correlation between the cosmological parameters, we include in the analysis the Planck data up to multipoles $\ell=650$ (as done in Ref.~\cite{ACT:2020gnv} to avoid the overlapping of the two experiments), and we study the combination ACT+Planck+BAO. As we can see in the last column of \autoref{tab1}, while all the other cosmological parameters are affected by the inclusion of the first peak from Planck, drifting away from the ACT only values, the 1$\sigma$ indication for a $\nu$DM coupling is strengthened as the errors shrink. In particular, we find $\log_{10}u_{\nu \rm{DM}}=-5.20^{+1.2}_{-0.74}$ at 68\% CL, with a definite distribution of the minimum of the $\chi2$ around this peak, as we can see in \autoref{fig:dist_chi2}, that is not disfavored by a Bayesian model comparison. It is important to notice here that this value of the coupling favored by the full combination is not in disagreement with the Planck data, as we can see in the top panel of \autoref{fig:residuals}, as in the Planck multipole range this model is indistinguishable from the $\Lambda$CDM scenario (see also the discussion in Ref.~\cite{Brax:2023rrf}). However, unfortunately, also this ACT+Planck+BAO dataset combination is not powerful enough to bound the interaction with a stronger statistical evidence, and this could certainly be a goal for future experiments. We will describe the phenomenological implications of having a $\nu$DM coupling different from zero in \autoref{sec:modelexample}. 

\begin{figure}
    \centering
    \includegraphics[width=\columnwidth]{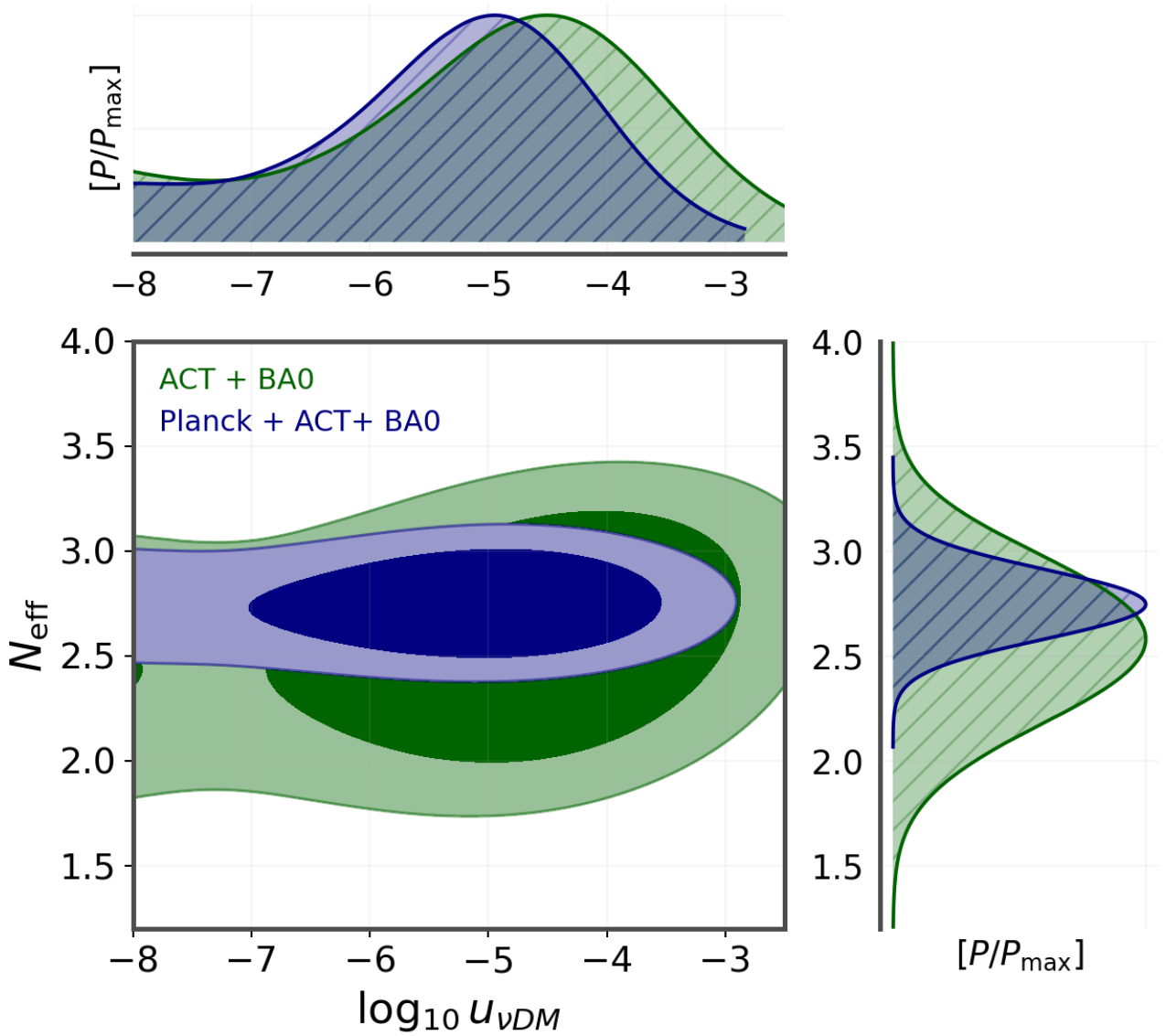}
    \caption{2D contours at 68\% and 95\% CL and 1D posterior probability distributions for the coupling parameter $\log_{10} u_{\nu \rm{DM}}$ and the effective number of relativistic degrees of freedom $N_{\rm eff}$, as obtained by the different combinations of CMB and BAO data listed in the legend.}
    \label{fig:2dneff}
\end{figure}

In a second step, we consider $N_{\rm eff}$ free to vary and report the constraints on the cosmological parameters at 68\% (and 95\% CL) in \autoref{tab2} for the temperature independent $\nu$DM cross-section case. The reason behind the variation of the effective number of relativistic degrees of freedom at recombination, different from the expected $N_{\rm eff} = 3.044$, is that we can expect a contribution to this parameter for the interaction between DM and neutrinos. However, as we can easily see in \autoref{tab2} and \autoref{fig:2dneff}, this is not the case for all the combinations of datasets explored in this work. Indeed, the constraints on $N_{\rm eff}$ obtained in this scenario are the same obtained assuming a model without coupling between DM and neutrinos, i.e. a $\Lambda$CDM+$N_{\rm eff}$ model. This happens because $N_{\rm eff}$ is very strongly constrained by the CMB data, and the preferred mean value is slightly below the expected $3.044$. However, interestingly, all the features regarding the results on the coupling in \autoref{tab1} are still valid in this one parameter extension. 

In particular, we find only an upper limit for $\log_{10}u_{\nu \rm{DM}}$ from Planck and Planck+BAO, while we have the same 1$\sigma$ indication for a non-vanishing $\nu$DM coupling by ACT as well as by its combination with BAO and the low-multipoles Planck data. We therefore obtain for the full dataset combination ACT+Planck+BAO data $\log_{10}u_{\nu \rm{DM}}=-5.29^{+1.3}_{-0.81}$ at 68\% CL ($\log_{10}u_{\nu \rm{DM}}<-4.21$ at 95\% CL). In the latter case the Bayes Factor weakly favors the interacting model against the model without interaction.

In conclusion, a non-zero coupling between DM and neutrinos has a very robust 1$\sigma$ indication coming from the ACT CMB data (alone and in combination with Planck and BAO) and it is also favored by a Bayesian model comparison. This remains true both fixing and varying $N_{\rm eff}$.

\section{Sterile neutrino portal to dark matter\label{sec:modelexample}}

We now illustrate the impact of the cosmological bounds we have found  on a specific model of DM-neutrino interactions. To this end, and to avoid bounds from the charged lepton sector, we focus on the scenario in which a fermionic DM species $\chi$ couple to a new scalar $\phi$ and a heavy neutral lepton $N$ mixing with the SM neutrinos, see also Ref.~\cite{Blennow:2019fhy} for a discussion on a model with  a mediator. The coupling in the dark sector reads $\mathcal{L} \supset -\,\phi\,\bar{\chi}\,(y_L\,N_L+y_R\,N_R) + \textrm{h.c.}$~\cite{Bertoni:2014mva,Batell:2017rol,Batell:2017cmf}. The Dirac fermion $N$ mixes with the SM neutrinos via a Yukawa-like coupling, $\mathcal{L} \supset -\lambda\,(\bar{L}\,\hat{H})\,N_R$, where $H$ is the Standard Model (SM) Higgs field, $\hat{H} = i\tau_2 H^\ast$, and $L$ is the lepton doublet. After electroweak symmetry breaking, this gives rise to the sterile neutrino mass eigenstate $\nu_4$ with admixtures of SM flavor eigenstates. We denote the relevant mixing angles by $U_{\ell 4}$, where $\ell = e,\mu,\tau$. In the following, we focus on the dominant mixing with the tau neutrinos and assume $U_{\tau 4} \neq 0 = U_{e4} = U_{\mu 4}$. This allows one to suppress bounds from the active-sterile neutrino mixing that are more prominent for the leptons from the first two generations~\cite{Abdullahi:2022jlv,Batell:2022xau}. For this reason, it is also useful to assume further that the sterile neutrino is more massive than the other dark sector species such that its dark decays $\nu_4\to \chi\phi$ are kinematically allowed while the semi-visible decay modes are highly suppressed. 

\begin{figure}[t]
    \centering
    \includegraphics[width=0.7\columnwidth]{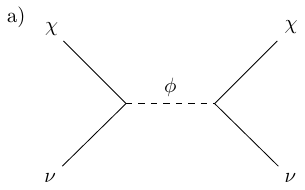}
    \includegraphics[width=0.7\columnwidth]{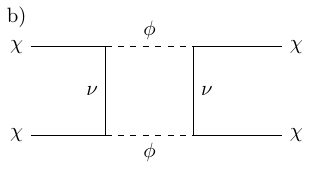}
\caption{Feynman diagrams for $\nu$DM interactions (top) as well as $\chi$ DM self-interactions (bottom) in the sterile neutrino portal model. Neutrino coupling to the dark species is suppressed by active-sterile neutrino mixing angle.}
    \label{fig:feynman}
\end{figure}

For the heavy dark scalar species and low neutrino energies, $m_\phi\gg m_\chi, E_\nu$, the $\nu$DM scattering cross section in this model is $T^2$-dependent and its present-day value reads~\cite{Bertoni:2014mva}
\begin{equation}
\label{eq:sterilesigmanuT2}
\sigma_{\chi\nu}\simeq (10^{-52}\,\textrm{cm}^2)\,\left(\frac{g}{0.1}\right)^4\,\left(\frac{100~\textrm{MeV}}{m_\phi}\right)^4\,\left(\frac{T}{T_0}\right)^2,
\end{equation}
where $g = y_L\,|U_{\tau 4}|$, $T_0 \simeq 0.2348~\textrm{meV}$~\cite{ParticleDataGroup:2022pth}, and we have assumed $\langle E_\nu^2\rangle = 15\,[\zeta(5)/\zeta(3)]\,T^2 \simeq 12.94\,T_\nu^2$ for the Fermi-Dirac statistics. The corresponding Feynman diagram is shown in \autoref{fig:feynman}. This predicted benchmark value of the cross section leads to $\log_{10}u_{\nu \rm{DM}}\ll-15$, and is orders of magnitude below the target values obtained in our cosmological simulations with $\sigma_{\nu\textrm{DM}}\sim T^2$ given in \autoref{tab3} and \autoref{tab4} in \hyperref[sec.T2]{Appendix A}. One could improve this by increasing the coupling constant $g$ close to the perturbativity limit and by reducing the DM mass $m_\chi$. The model, however, is then also subject to various constraints. Strong positive temperature-dependence of the $\nu$DM cross section is bounded from above by an expected attenuation of high-energy neutrino flux from distant blazars~\cite{Cline:2022qld,Ferrer:2022kei,Cline:2023tkp}. In the scenario discussed here, though, $\sigma_{\nu\textrm{DM}}$ becomes suppressed for $E_\nu\sim 100~\textrm{TeV}$ relevant for these bounds so that they are alleviated. A similar conclusion  is true for the constraints based on the expected anisotropy of high-energy extragalactic neutrinos interacting in the Galactic DM halo~\cite{Arguelles:2017atb,Pandey:2018wvh}. Important bounds, however, can be deduced from observables sensitive to lower neutrino energies where the $T^2$ dependence of the cross section holds. In particular, Lyman-$\alpha$ observations can give rise to substantially stronger limits on $\sigma_{\nu\textrm{DM}}$ in this case than the ones derived from the CMB data~\cite{Mangano:2006mp,Wilkinson:2013kia,Wilkinson:2014ksa,Escudero:2015yka}. The $\nu$DM interactions of this type can also be constrained by observing dips in the diffuse supernova neutrino background. This affects scenarios with $g\gtrsim 0.1$ and the dark sector masses of the order of tens of MeV or below~\cite{Farzan:2014gza}. See also Refs.~\cite{Kelly:2018tyg,Alvey:2019jzx,Choi:2019ixb,Jho:2021rmn,Ghosh:2021vkt,Lin:2022dbl} for further astrophysical bounds on $\nu$DM interactions. We conclude that it remains hard to reconcile this scenario in the limit of $m_\phi\gg m_\chi,E_\nu$ with our cosmological fits.

The specific temperature-independent regime of the cross section can also be obtained in this model at low energies assuming a mass degeneracy between the dark sector fermion and scalar species, $m_\chi\simeq m_\phi$~\cite{Brax:2023rrf}. In the limit of a small mass splitting, $(m_\phi-m_\chi)\ll E_\nu$, the $\nu$DM cross section in this scenario is given by
\begin{eqnarray}
\label{eq:sigmachinuT0}
\sigma_{\chi\nu} & \simeq & (10^{-34}\,\textrm{cm}^2)\,\left(\frac{g}{0.01}\right)^4\,\left(\frac{20~\textrm{MeV}}{m_{\chi}}\right)^2\\
& &\times\left[1+0.075\,\left(\frac{m_\chi}{20~\textrm{MeV}}\right)^2\,\left(\frac{T_{\textrm{rec.}}}{T_\nu}\right)^2\left(\frac{\delta}{10^{-8}}\right)^2\right],\nonumber
\end{eqnarray}
where $\delta = (m_\phi-m_\chi)/m_\chi$. As can be seen, in order to guarantee that $\sigma_{\chi\nu}$ becomes effectively temperature-independent until the recombination epoch with $T_{\textrm{rec.}}\sim 0.25~\textrm{eV}$, one requires $(m_\phi-m_\chi)\lesssim 100~\textrm{meV}$ for the dark sector masses of the order of tens of MeV. For larger mass splittings, the cross section first starts to grow with a decreasing temperature, $\sigma_{\nu\textrm{DM}}\sim T^{-2}$, in a narrow region relevant for $m_\phi-m_\chi \lesssim E_\nu$, and then it enters the $\sigma_{\nu\textrm{DM}}\sim T^2$ regime characteristic of  many $\nu$DM interaction models with a substantial mass difference between the dark sector particles. The transition between the two regimes occurs via a possible resonant $\phi$ production for which the $\nu$DM cross section substantially grows. 

\begin{figure}[t]
    \centering
    \includegraphics[width=\columnwidth]{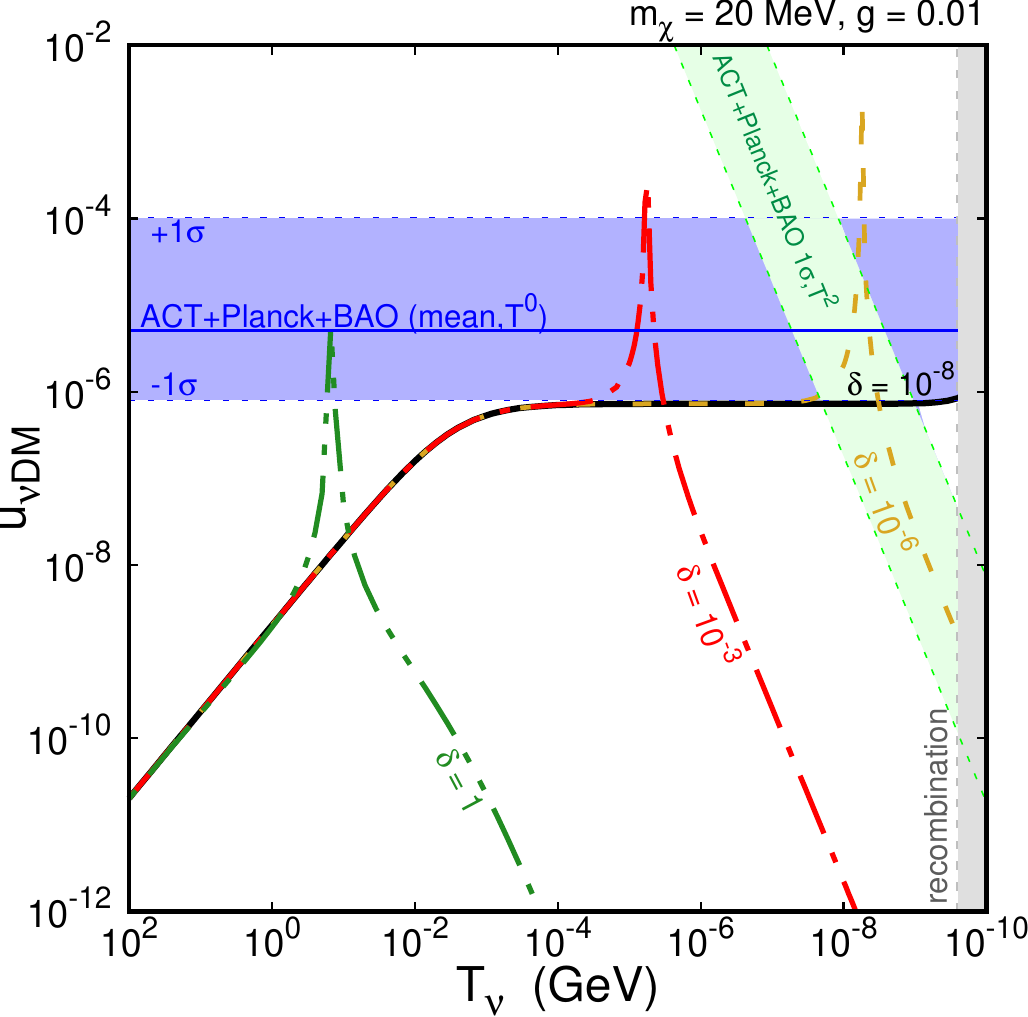}
\caption{Dark matter scattering cross section off neutrinos in the sterile neutrino portal model shown in terms of the $u_{\nu\textrm{DM}}$ parameter as a function of the temperature for several benchmark scenarios with $m_\chi = 20~\textrm{MeV}$, $g=0.01$, and $\delta = 10^{-8}, 10^{-6}, 10^{-3}, 1$, where $\delta = (m_\phi-m_\chi)/m_\chi$. We also show $1\sigma$ regions around the mean fitted value of $u_{\nu\textrm{DM}}$ in the $T^0$ and $T^2$ regimes of the $\nu$DM cross section as blue- and green-shaded regions, respectively. The recombination epoch is indicated with a gray-shaded region on the right.}
    \label{fig:sterileneutrinocrosssection}
\end{figure} 

We illustrate the temperature dependence of the $u_{\nu\textrm{DM}}$ parameter obtained in this model in \autoref{fig:sterileneutrinocrosssection} for several benchmark scenarios with $\delta = 10^{-8},10^{-6},10^{-3},1$ and $m_\chi = 20~\textrm{MeV}$, $g=0.01$. In the plot, we show with a black solid line the expected result for the model with a tiny $\delta\sim 10^{-8}$. As can be seen, in this case, the $\nu$DM cross section is effectively temperature-independent for $\textrm{MeV}\gtrsim T \gtrsim T_{\textrm{rec.}}$ and it can fit the $1\sigma$ region around the mean value of our ACT+Planck+BAO fit shown as a blue-shaded horizontal region. Instead, for larger temperatures and $E_\nu\gtrsim m_\chi$ the cross section becomes suppressed, $\sigma_{\nu\textrm{DM}}\sim 1/E_\nu$. In this case, the transition between the $T^0$ and $T^2$ regimes occurs for $T<T_{\textrm{rec.}}$  This transition is shifted towards larger temperatures for  increasing $\delta$'s. Interestingly, for $\delta = 10^{-6}$, the $\nu$DM cross section can fit the ACT+Planck+BAO $1\sigma$ regions characteristic for both the $T^0$ regime at $T\gtrsim 10~\textrm{eV}$ and the effectively $T^2$-dependent regime characteristic for $T\sim \textrm{eV}$, i.e. around the time of matter-radiation equality. The latter is shown as green-shaded region in the plot following \autoref{tab4}. Instead, for $\delta\gtrsim 10^{-5}$ the scattering cross section becomes much suppressed at lower temperatures, although it can still be within the ACT+Planck+BAO $1\sigma$ region for higher $T$ during the radiation-dominated epoch unless the mass splitting grows too high, $\delta\gtrsim 0.1$.

We stress that a low mass splitting of the order of $\delta \sim 10^{-8}$ requires some fine-tuning of the model parameters and could be  radiatively unstable. In particular, corrections to the $\phi$ mass can occur at a loop level via a radiatively generated $\lambda_{\phi H}|\phi|^2|H|^2$ term, where one estimates $\lambda_{\phi H}\sim (y_L^2\lambda^2/16\pi^2)\,\log(\Lambda_{\textrm{UV}}/m_N^2) = (g^2/16\pi^2)\,(m_N/v)^2\,\log(\Lambda_{\textrm{UV}}/m_N^2)$~\cite{Batell:2017cmf}. In the second step we used $\lambda = U_{\ell 4}(m_N/v)$ and introduced the cutoff scale $\Lambda_{\textrm{UV}}$. This can lead to $\textrm{keV}-\textrm{MeV}$ corrections to $m_\phi$ in the region of the parameter space of our model which should be taken into account when studying the mass-degenerate regime. These corrections can be further modified by introducing an additional explicit $\lambda_{\phi H}^{\prime}|\phi|^2|H|^2$ interaction term which is not forbidden in the model under study. In the following, we assume a strong mass degeneracy between the $\chi$ and $\phi$ dark sector species to illustrate the interesting phenomenology of the model while keeping agnostic about a possible origin of such a degeneracy.

\begin{figure}[t]
    \centering
    \includegraphics[width=\columnwidth]{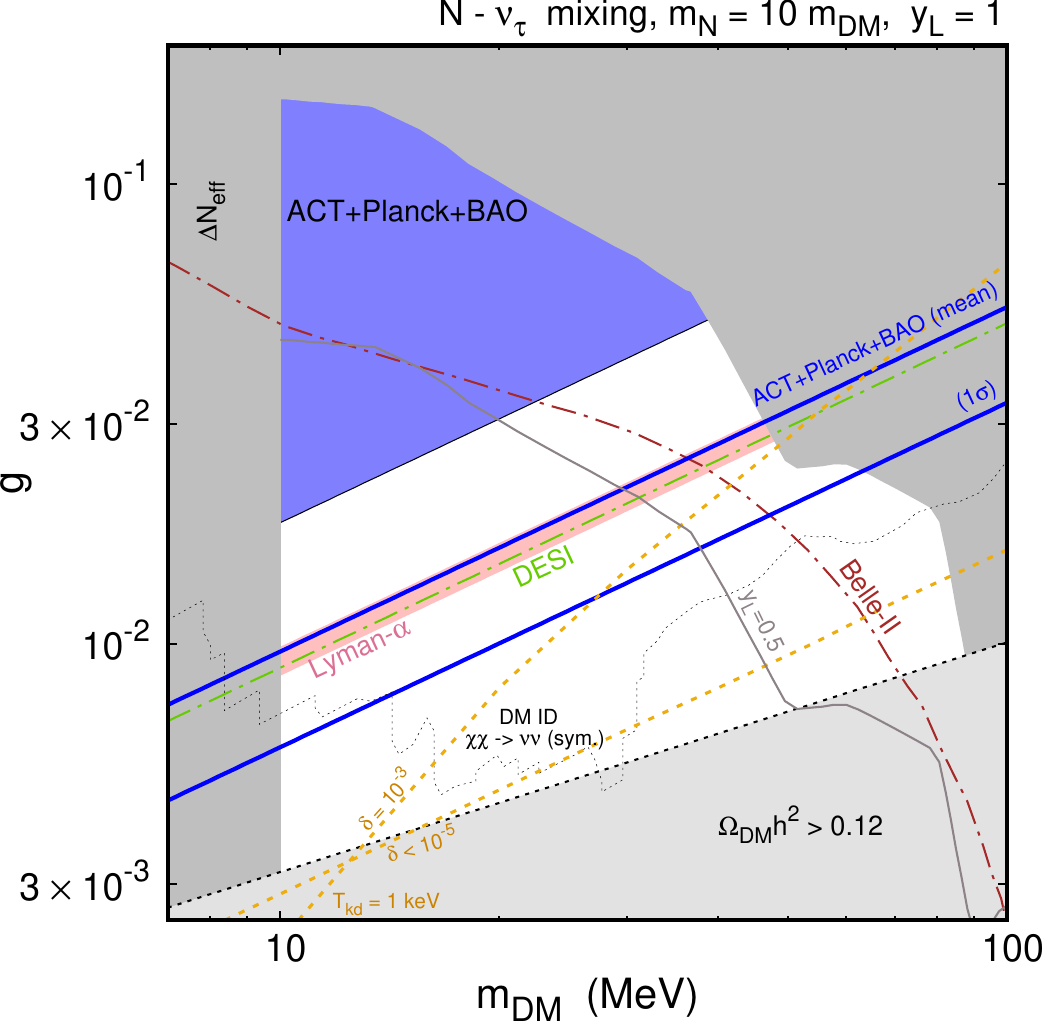}
    \caption{The parameter space of the neutrino portal DM model shown in the $(m_{\textrm{DM}},g)$ plane. One assumes $m_N = 10~m_{\textrm{DM}}$, $y_L = 1$ and the mass-degenerate scenario with $m_{\textrm{DM}}\equiv m_\chi\simeq m_\phi$. ACT+Planck+BAO exclusion bounds are shown as a blue-shaded region, while the relevant average value of the $\log_{10}u_{\nu \rm{DM}}$ parameter obtained in the fitting, and its $1\sigma$ deviation below the mean are illustrated with solid blue lines. Constraints on sterile-active neutrino mixing, the effective number of relativistic degrees of freedom $\Delta N_{\textrm{eff}}$, and $\chi$ DM relic density are shown as grey-shaded regions. For comparison, we also present such bounds derived for a lower value of the Yukawa parameter $y_L=0.5$ as indicated with a gray solid line. Lyman-$\alpha$ best-fit region is shown with red-shaded color~\cite{Hooper:2021rjc}. The $\nu$DM kinetic decoupling occurs at $T_{kd}\simeq 1~\textrm{keV}$ for $\delta=10^{-3}$ and $10^{-8}$ along orange dotted lines, where $\delta = (m_\phi-m_\chi)/m_\chi$. DM indirect detection constraint on present-day annihilations of the symmetric $\chi$ DM component is shown with a black dotted line. This bound is avoided in the asymmetric DM regime. Future expected sensitivity of the Belle-II~\cite{Kobach:2014hea} and DESI~\cite{Escudero:2015yka} experiments are shown with red and light-green dash-dotted lines, respectively.}
    \label{fig:sterileneutrinosensitivity}
\end{figure}

An example of  allowed region in the parameter space of the model is shown in \autoref{fig:sterileneutrinosensitivity}. In the plot, we assume $m_{\nu_4} = 10\,m_\chi$, $y_L=1$, and $\delta\sim 10^{-8}$. Current bounds from active-sterile neutrino mixing~\cite{Batell:2017cmf,Cvetic:2017gkt,BaBar:2022cqj} and BBN constraints due to a $\chi$ DM contribution to the number of relativistic degrees of freedom $\Delta N_{\textrm{eff}}$~\cite{Boehm:2013jpa} are shown as gray-shaded regions, cf. Ref.~\cite{Brax:2023rrf} for further discussion. These bounds on the coupling $g$ become more severe for lower values of the Yukawa-like coupling $y_L$ as it is shown with the gray solid line in the plot which corresponds to $y_L=0.5$. They also grow with the increasing mass of the heavy neutral lepton $N$. The mean value of the $\nu$DM scattering cross section, which fits the ACT+Planck+BAO data as given in \autoref{tab2}, and the relevant $1\sigma$ downward deviation are shown with solid blue lines, as indicated in the plot. Strikingly, a corresponding $1\sigma$ range of non-zero preferred values of the $\nu$DM interaction cross section has recently been found in the Lyman-$\alpha$ data~\cite{Hooper:2021rjc}, as shown with red-shaded color. In the plot, we also show future expected bounds from the Dark Energy Spectroscopic Instrument (DESI)~\cite{Escudero:2015yka} and Belle-II experiment~\cite{Kobach:2014hea}. As can be seen, cosmological fits to the CMB data can be obtained in the region of the parameter space of the model under study which is currently not excluded by accelerator-based searches but remains within the reach of such experiments and future cosmological surveys~\cite{Abazajian:2019eic,SimonsObservatory:2018koc,NASAPICO:2019thw,CMB-HD:2022bsz}.

We also stress that the simplest version of the model with a single-component DM interacting strongly with neutrinos might suffer from other important astrophysical and cosmological bounds. In particular, interactions between the DM species and neutrinos can also impact small-structure of the Universe due to late kinetic decoupling of $\nu$DM interactions~\cite{Shoemaker:2013tda,Boehm:2014vja,Bertoni:2014mva,Schewtschenko:2015rno}. This can suppress the growth of small-scale structures via non-negligible neutrino pressure affecting DM density perturbations. The kinetic decoupling temperature of such interactions should remain of the order of $T_{kd}\sim \textrm{keV}$ to avoid too strong a suppression while one could then also potentially address the persisting missing satellite problem within the framework of this model~\cite{Bertoni:2014mva}. We estimate the kinetic decoupling temperature by comparing the $\nu$DM interaction rate that sizeably change DM momentum to the Hubble expansion rate of the Universe, $\gamma(T_{kd}) = H(T_{kd})$, following Ref.~\cite{Gondolo:2012vh}. In the strong mass-degenerate regime, one obtains 
\begin{equation}
T_{kd}\Big|_{m_\phi\simeq m_\chi}\simeq (0.12~\textrm{keV})\,\left(\frac{0.01}{g}\right)^2\,\left(\frac{m_\chi}{20~\textrm{MeV}}\right)^{3/2}.
\end{equation}
For this value of the kinetic decoupling temperature, $T_{kd}\sim 0.1~\textrm{keV}$, acoustic oscillations could erase primordial density fluctuations up to a large cutoff scale, $M_{\textrm{cutoff}}\sim 10^{11}\,M_{\odot}\,(0.1~\textrm{keV}/T_{kd})^3$~\cite{Loeb:2005pm} where $M_{\odot}$ is the solar mass.\footnote{Notably, the impact of this on our cosmological fits presented in \autoref{sec:results} is expected to be negligible, as it affects large wavenumbers with $k_{\textrm{cut}}\sim 3.7~\textrm{Mpc}^{-1}$ which translates into $\ell>50$k in the CMB power spectrum.}

In order to avoid such a large suppression of the structure growth, departures from the benchmark scenario presented in \autoref{fig:sterileneutrinosensitivity} can be considered. In particular, only a fraction of DM could be strongly-interacting with neutrinos while the remaining part could undergo earlier kinetic decoupling. The impact of such a two-component DM scenario on neutrino free streaming can be kept equally significant by appropriately increasing the coupling constant $g$ of the interacting component. Lower values of $M_{\textrm{cutoff}}$ can also be obtained for larger mass splittings between the $\chi$ and $\phi$ species. This is particularly the case for the scenarios predicting the transition from the $T^0$ to the $T^2$ regime at temperatures $T\gtrsim \textrm{keV}$. We illustrate this in \autoref{fig:sterileneutrinosensitivity} with orange dotted lines along which one predicts $T_{kd}=1~\textrm{keV}$ for $\delta = 10^{-8}$ and $10^{-3}$. The difference between the shape of the two lines indicates the distinction between the kinetic decoupling occurring in the $T^0$ and $T^2$ regimes of the $\nu$DM cross section, respectively. In the latter case, $\sigma_{\chi\nu}\sim 1/m_{\chi}^4$ and its strong increase for decreasing $m_\chi$ needs to be compensated for with smaller values of the coupling constant $g$, cf. \autoref{eq:sterilesigmanuT2}. As can be seen, in the $\delta=10^{-3}$ case, the $T_{kd}=1~\textrm{keV}$ line for $m_\chi\gtrsim 30~\textrm{MeV}$ corresponds to the $1\sigma$ ACT+Planck+BAO region around the mean value of $\bar{u}_{\nu\textrm{DM}}$. We leave detailed analyses of the two-component DM model and the models with a larger mass splitting $\delta$ characterized by earlier $\nu$DM kinetic decoupling, for future studies.

Small mass splitting between $\chi$ and $\phi$ can also be constrained from possible $\chi$ DM self interactions. In particular, fermion upscattering to on-shell scalars with the exchange of Majorana SM neutrinos can be made kinematically available, $\chi\chi\to\phi\phi$, with the relevant cross section in the Born approximation given by $\sigma_{\chi\chi\to\phi\phi}/m_\chi \simeq (5\times 10^{-9}\,\textrm{cm}^2/\textrm{g})\,(g/0.01)^4\,(10~\textrm{MeV}/m_\chi)^3$. Here, we assume $m_\chi\,v\gg m_\nu$ and $v\gtrsim 2\sqrt{\delta}$. The latter condition is to ensure that $\chi$ species have enough kinetic energy to produce two on-shell scalars $\phi$. Assuming $\delta\sim 10^{-8}$, this can be effective already for $v\sim \textrm{tens of km/s}$, i.e., at the Galactic scale, while for $\delta\gtrsim 10^{-5}$ one requires $v\sim (\textrm{few}\times 10^3\,\textrm{km/s})$ and the upscattering can only happen at scales of galaxy clusters. Inelastic $\chi$ DM self-scattering followed by a decay $\phi\to\chi\nu$ can lead to, e.g., the effective cooling of the inner DM core when final-state neutrinos escape this region. As a result, a central DM density slope can be increased, cf. Ref.~\cite{Das:2017fyl} for discussion about this effect and further observational consequences. The value of the cross section mentioned above is lower than current bounds on DM self-scatterings, so such effects are expected to be very mild. However, it is important to note that further corrections can arise from multiple neutrino exchange diagrams. This can lead to sizeable effects for loop-induced $\chi\chi\to\chi\chi$ self-interactions via off-shell scalars $\phi$~\cite{Orlofsky:2021mmy}, see \autoref{fig:feynman} for the corresponding box diagram. In the limit of the effective contact operator, when $\phi$ scalars can be integrated out for interactions of non-relativistic $\chi$s, this leads to a long-range repulsive potential $V\sim 1/r^5$. Bounds at the level of $g\lesssim \mathcal{O}(0.1)$ for $m_\chi\simeq 20~\textrm{MeV}$ and $\delta\sim 0.1$ can then be derived from too strong  $\chi$ DM self scatterings. These constraints, however, are expected to be modified when going beyond the regime of the contact operator in the strongly mass-degenerate limit, see also Ref.~\cite{Xu:2021daf} for the discussion. Notably, DM self-interaction bounds can also be weakened assuming that $\chi$ corresponds to only a fraction of DM. Finally, neutrino self-interactions can also be induced at a loop level with $\chi$ and $\phi$ species exchanged in a box diagram. The corresponding effective coupling in the four-neutrino contact operator, $G_{\nu\textrm{SI}}\,(\bar{\nu}\gamma^\mu P_L\nu)\,(\bar{\nu}\gamma^\mu P_L\nu)$, is given by $G_{\nu\textrm{SI}}\sim (10^{-3}\,G_F)\,(g/0.01)^4\,(20\,\textrm{MeV}/m_\chi)^2$~\cite{Orlofsky:2021mmy}. This, however, remains much below current constraints, cf. Refs.~\cite{Deppisch:2020sqh,Berryman:2022hds}.

Last but not least, we comment on the $\chi$ DM relic density. In the mass-degenerate regime, the $\chi$ abundance is dictated by both $\chi$ and $\phi$ annihilations into SM neutrinos, $(\chi\bar{\chi}/\phi\bar{\phi})\to\nu\bar{\nu}$ with the latter followed by the decay, $\phi\to \chi\nu$. The corresponding lifetime reads
\begin{equation}
\tau_\phi \simeq (0.1~\textrm{sec})\,\left(\frac{0.01}{g}\right)^2\,\left(\frac{20~\textrm{MeV}}{m_\phi}\right)\,\left(\frac{10^{-8}}{\delta}\right).
\end{equation}
For the mass splitting $\delta\gtrsim 10^{-8}$ and other parameters corresponding to the benchmark in \autoref{eq:sigmachinuT0}, the $\phi$ species decay early, before the Big Bang Nucleosynthesis (BBN) epoch, so they do not constitute DM. We show the thermal target for $\chi$ DM with a black dotted line in \autoref{fig:sterileneutrinosensitivity}. Importantly, the annihilation cross section corresponding to the benchmark values of $\sigma_{\chi\nu}$ from the ACT+Planck+BAO fit exceeds this thermal target. This leads to a suppressed $\chi$ thermal relic abundance. The correct value of $\Omega_\chi h^2$ can, instead, be obtained as a result of the $\chi$ freeze-out in the presence of initial asymmetry between $\chi$ and $\bar{\chi}$~\cite{Graesser:2011wi,Iminniyaz:2011yp}. While an additional contribution to the $\chi$ relic density could also result from non-thermal processes, it remains essential to suppress the symmetric DM component also in this case. Otherwise, stringent DM indirect detection (ID) bounds appear from $\chi\bar{\chi}\to\nu\bar{\nu}$ annihilations, cf. Ref.~\cite{Arguelles:2019ouk} for review. We show such current bounds in \autoref{fig:sterileneutrinosensitivity} based on data from the KamLAND~\cite{KamLAND:2011bnd} and Super-Kamiokande collaborations~\cite{Super-Kamiokande:2015qek,Olivares-DelCampo:2017feq}. These bounds do not apply to asymmetric DM scenario. They could also be weakened if $\chi$ corresponds to only a fraction of the total DM relic density, in which case the annihilation rate of the $\chi$ DM component remains suppressed by a factor $(\Omega_\chi/\Omega_{\textrm{total DM}})^2$. Instead, direct detection bounds on $\chi$ DM remain weak as the relevant couplings to quarks arise only at the loop level~\cite{Batell:2017cmf}.

\section{Conclusions} \label{sec:concl}

In a recent study~\cite{Brax:2023rrf}, we have pointed out that precise measurements of CMB anisotropies on small angular scales (i.e., large multipoles $l\gtrsim 3000$) can provide crucial insights into neutrino DM interactions. Specifically, we have demonstrated that accurate measurements with a few per cent uncertainty on those scales can yield a significant amount of information, whereas detecting these interactions on larger angular scales (such as those probed by the Planck experiment) would require substantially greater sensitivity. 

Interestingly, although the most recent observations of the cosmic microwave background released by the Planck satellite do not provide any concrete evidence in support of the aforementioned models, the analysis of small-scale CMB data from the Atacama Cosmology Telescope shows a preference for a non-zero interaction strength, underscoring the significance of current, forthcoming and future high-multipole measurements to  constrain better such scenarios.

In order to validate further  the robustness of these findings, in this work we have presented an extended and comprehensive analysis of $\nu$DM interactions in the cosmic microwave background, conducting a significant number of additional tests.

First and foremost, we have shown that our results remain robust when combining observations from the two most accurate CMB experiments to date (Planck and ACT), both including and excluding astrophysical measurements of Baryon Acoustic Oscillations and Redshift Space Distortions. In presence of the small-scale CMB measurements provided by ACT, all of these combinations of independent datasets confirm the same preference for a non-zero interaction strength, see also \autoref{tab1}.

Secondly, we have proved that the same preference is found by both fixing the effective number of relativistic particles to the Standard Model value of $N_{\rm eff}=3.044$ and considering it as a free parameter of the cosmological model, see \autoref{tab2} and \autoref{fig:2dneff}. 

To get further insights and to consolidate our results, we have conducted a thorough examination of the data provided by both CMB experiments that are summarized in \autoref{fig:residuals}. We have verified that the peak in the distribution of the interaction strength is associated with a genuine and significant reduction of the $\chi^2$ of the fit, as also evident from \autoref{fig:dist_chi2}. These two figures support the conclusion that the observed preference for $\nu$DM interactions is not an artifact of the sampling method, nor a volume effect, but instead an actual preference of the ACT data.

In this respect, although such a preference for a non-zero interaction strength is not directly supported by Planck, it is noteworthy that the two experiment do not conflict with each other about the predicted value for this parameter: our analysis suggest that the ACT’s indication for a non-zero coupling can be easily explained by the larger effects of couplings of the order of $u_{\nu \textrm{DM}}\sim 10^{-6} - 10^{-4}$ in the multipole range probed by this experiment, see also \autoref{fig:theory}. In addition, we have assessed the plausibility of both interacting and non-interacting models in explaining the current observations by conducting a Bayesian model comparison for \textit{all} the different combinations of data and models studied in this paper. We have found that, while both models are plausible, the interacting $\nu$DM (+$N_{\rm eff}$) scenario is often preferred over a baseline $\Lambda$CDM (+$N_{\rm eff}$) cosmology; even considering the Planck data alone. Therefore we can conclude that Planck does not conflict with the possibility of an interacting $\nu$DM cosmology.

While our analysis primarily focuses on the case of a temperature-independent cross-section, in \hyperref[sec.T2]{Appendix A} we also consider the possibility of a temperature-dependent cross-section $\sigma_{\nu\textrm{DM}} \propto T^2$. In this case, the results obtained fixing $N_{\rm eff}$ are summarized in \autoref{tab3}, while \autoref{tab4} summarizes those obtained relaxing this relation and considering $N_{\rm eff}$ as a free parameter of the sample. Interestingly, a similar $\sim 1 \sigma$ preference for a non-zero interaction strength emerges in both cases. However, due to the smaller effects on the CMB anisotropies predicted for the temperature dependent case, such a preference is somewhat mitigated compared to the temperature-independent scenario, see also \hyperref[sec.Tplot]{Appendix B} where the correlation between cosmological parameters is shown for all the cases under study in this work. Therefore, we advise caution when interpreting the results for the temperature-dependent cross-section and suggest that this case warrants a more detailed and focused analysis.

Finally, we have provided a detailed example showing how a sterile neutrino portal between DM and the Standard Model could accommodate such a coupling. We show that the preferred values of the $\nu$DM cross section from our cosmological fits can be made consistent with the current bounds on light right-handed neutrinos mixing with the SM neutrinos and constraints from interactions of neutrinos produced in astrophysical sources while they remain within the reach of future searches. Notably, substantial $\nu$DM couplings can have important consequences for structure formation both by  affecting DM density perturbations due to late kinetic decoupling and due to substantial DM self-interactions at late times. Reconciling this scenario with all the observations might require going beyond the simple single-component DM model with a temperature-independent cross section up to the recombination epoch. We leave a detailed exploration of such scenarios for future studies investigating possible intriguing connections between the elusive SM neutrinos and DM.

\begin{acknowledgments}
CvdB is supported (in part) by the Lancaster–Manchester–Sheffield Consortium for Fundamental Physics under STFC grant: ST/T001038/1. EDV is supported by a Royal Society Dorothy Hodgkin Research Fellowship. ST is supported by the grant ``AstroCeNT: Particle Astrophysics Science and Technology Centre" carried out within the International Research Agendas programme of the Foundation for Polish Science financed by the European Union under the European Regional Development Fund. ST is also supported in part by the National Science Centre, Poland, research grant No. 2021/42/E/ST2/00031. ST acknowledges the support of the Institut Pascal at Université Paris-Saclay during the Paris-Saclay Astroparticle Symposium 2022, with the support of the P2IO Laboratory of Excellence (program “Investissements d’avenir” ANR-11-IDEX-0003-01 Paris-Saclay and ANR-10-LABX-0038), the P2I axis of the Graduate School Physics of Université Paris-Saclay, as well as IJCLab, CEA, IPhT, APPEC, the IN2P3 master projet UCMN and ANR-11-IDEX-0003-01 Paris-Saclay and ANR-10-LABX-0038. This article is based upon work from COST Action CA21136 Addressing observational tensions in cosmology with systematics and fundamental physics (CosmoVerse) supported by COST (European Cooperation in Science and Technology). We acknowledge IT Services at The University of Sheffield for the provision of services for High Performance Computing.
\end{acknowledgments}

\appendix

\section{$T^2$ Cross-Section}
\label{sec.T2}

\begin{table*}[htbp!]
\begin{center}
\renewcommand{\arraystretch}{1.5}
\resizebox{\textwidth}{!}{
\begin{tabular}{l c c c c c c c c c c c c c c c }
\hline
\textbf{Parameter} & \textbf{ Planck } & \textbf{ Planck + BAO } & \textbf{ ACT } & \textbf{ ACT + BAO } & \textbf{ ACT + Planck + BAO } \\ 
\hline\hline
$ \Omega_\mathrm{b} h^2  $ & $  0.02239\pm 0.00015 $ & $  0.02239\pm 0.00014 $ & $  0.02151\pm 0.00032 $ & $  0.02148\pm 0.00030 $ & $0.02235\pm 0.00012$\\ 
$ \Omega_\mathrm{c}^{\nu\mathrm{DM}} h^2  $ & $  0.1195\pm 0.0012 $ & $  0.11950\pm 0.00094 $ & $  0.1173\pm 0.0039 $ & $  0.1196\pm 0.0015 $ & $0.11973\pm 0.00097$ \\ 
$ 100\theta_\mathrm{s}  $ & $  1.04189\pm 0.00029 $ & $  1.04188\pm 0.00029 $ & $  1.04342\pm 0.00072 $ & $  1.04321\pm 0.00064 $ & $1.04202\pm 0.00027$ \\ 
$ \tau_\mathrm{reio}  $ & $  0.0535\pm 0.0076 $ & $  0.0529\pm 0.0070 $ & $  0.063\pm 0.015 $ & $  0.058\pm 0.013 $ & $0.0553\pm 0.0065$ \\ 
$ \log(10^{10} A_\mathrm{s})  $ & $  3.041\pm 0.015 $ & $  3.040\pm 0.014 $ & $  3.046\pm 0.031 $ & $  3.040\pm 0.029 $ & $3.051\pm 0.013$ \\ 
$ n_\mathrm{s}  $ & $  0.9654\pm 0.0042 $ & $  0.9654\pm 0.0036 $ & $  1.007\pm 0.016 $ & $  1.002\pm 0.013 $ & $0.9678\pm 0.0035$ \\ 
$ log_{10}u_{\nu \rm{DM}}  $ & $  <-15.4\, (< -14.1 ) $ & $  <-15.35\, (< -14.3 ) $ & $  -15.2^{+1.8}_{-1.1}\, (< -13.9 ) $ & $  -15.3^{+1.8}_{-1.1}\, (< -13.9 ) $ & $-14.9^{+1.5}_{-0.63} (< -14.3)$ \\ 
$ H_0  $ & $  68.06\pm 0.55\, ( 68.1^{+1.1}_{-1.1} ) $ & $  68.06\pm 0.42\, ( 68.06^{+0.82}_{-0.86} ) $ & $  68.6\pm 1.6\, ( 68.6^{+3.3}_{-3.1} ) $ & $  67.68\pm 0.58\, ( 67.7^{+1.1}_{-1.1} ) $ & $67.99\pm 0.42$ \\ 
$ \sigma_8  $ & $  0.8212\pm 0.0062\, ( 0.821^{+0.012}_{-0.012} ) $ & $  0.8206\pm 0.0059\, ( 0.821^{+0.011}_{-0.011} ) $ & $  0.834\pm 0.017\, ( 0.834^{+0.032}_{-0.034} ) $ & $  0.838\pm 0.012\, ( 0.838^{+0.023}_{-0.023} ) $ & $0.8269\pm 0.0061$ \\ 
\hline
%$\Delta \chi^2$ & $ -9.77 $ & $ -9.07 $ & $ -3.21 $ & $ -3.11 $ & $ -6.47 $ \\
$\ln BF$ & $ -5.54 $ & $ -4.3 $ & $ 0.0374 $ & $ 0.421 $ & $ 0.398 $ \\
\hline \hline
\end{tabular} }
\end{center}
 \caption{\textbf{$T^2$ cross section:} We report the 68\% (95\%) CL constraints/bounds on the cosmological parameters above the line, while below the line we have the improvement of the $\chi^2$ of the best fit and the Bayes Factor, with respect to the $\Lambda$CDM scenario.}
\label{tab3}
\end{table*}

\begin{table*}
\begin{center}
\renewcommand{\arraystretch}{1.5}
\resizebox{\textwidth}{!}{
\begin{tabular}{l c c c c c c c c c c c c c c c }
\hline
\textbf{Parameter} & \textbf{ Planck } & \textbf{ Planck + BAO } & \textbf{ ACT } & \textbf{ ACT + BAO } & \textbf{ ACT + Planck + BAO } \\ 
\hline\hline
$ \Omega_\mathrm{b} h^2  $ & $  0.02228\pm 0.00022 $ & $  0.02230\pm 0.00019 $ & $  0.02109\pm 0.00045 $ & $  0.02106\pm 0.00038 $ & $0.02210\pm 0.00017$\\ 
$ \Omega_\mathrm{c}^{\nu\mathrm{DM}} h^2  $ & $  0.1177\pm 0.0030 $ & $  0.1176\pm 0.0029 $ & $  0.1105\pm 0.0065 $ & $  0.1086\pm 0.0058 $ & $0.1147\pm 0.0024$\\ 
$ 100\theta_\mathrm{s}  $ & $  1.04219\pm 0.00051 $ & $  1.04219\pm 0.00050 $ & $  1.0445\pm 0.0012 $ & $  1.0448\pm 0.0011 $ & $1.04279\pm 0.00043$\\ 
$ \tau_\mathrm{reio}  $ & $  0.0518\pm 0.0074 $ & $  0.0526\pm 0.0071 $ & $  0.060\pm 0.015 $ & $  0.061\pm 0.013 $ & $0.0547\pm 0.0067$\\ 
$ \log(10^{10} A_\mathrm{s})  $ & $  3.033\pm 0.017 $ & $  3.034\pm 0.016 $ & $  3.023\pm 0.037 $ & $  3.020\pm 0.030 $ & $3.035\pm 0.016$\\ 
$ n_\mathrm{s}  $ & $  0.9601\pm 0.0085 $ & $  0.9612\pm 0.0070 $ & $  0.969\pm 0.033 $ & $  0.969\pm 0.023 $ & $0.9568\pm 0.0066$\\ 
$ N_\mathrm{eff}  $ & $  2.91\pm 0.19\, ( 2.91^{+0.38}_{-0.37} ) $ & $  2.93\pm 0.17\, ( 2.93^{+0.33}_{-0.35} ) $ & $  2.49\pm 0.44\, ( 2.49^{+0.87}_{-0.83} ) $ & $  2.43\pm 0.33\, ( 2.43^{+0.69}_{-0.67} ) $ & $ 2.73\pm 0.14 \, (2.73^{+0.30}_{-0.30})$\\ 
$ log_{10}u_{\nu \rm{DM}}  $ & $ < -15.4\, (< -14.1 ) $ & $  < -15.35\, (< -14.0 ) $ & $  -15.2^{+1.7}_{-1.2}\, (< -13.8 ) $ & $  -15.3^{+1.6}_{-1.3}\, (< -13.8 ) $ & $-15.1^{+1.7}_{-0.90} \, (< -13.8)$\\ 
$ H_0  $ & $  67.2\pm 1.4\, ( 67.2^{+2.8}_{-2.7} ) $ & $  67.3\pm 1.1\, ( 67.3^{+2.1}_{-2.2} ) $ & $  64.3\pm 3.6\, ( 64.3^{+7.0}_{-7.0} ) $ & $  64.4\pm 1.9\, ( 64.4^{+3.9}_{-3.6} ) $ & $66.1\pm 1.0\, (66.1^{+2.1}_{-2.0})$\\ 
$ \sigma_8  $ & $  0.815\pm 0.010\, ( 0.815^{+0.020}_{-0.020} ) $ & $  0.8151\pm 0.0097\, ( 0.815^{+0.018}_{-0.019} ) $ & $  0.810\pm 0.025\, ( 0.810^{+0.050}_{-0.047} ) $ & $  0.804\pm 0.021\, ( 0.804^{+0.042}_{-0.040} ) $ & $0.8116\pm 0.0094 \, (0.812^{+0.019}_{-0.018})$\\ 
\hline
%$\Delta \chi^2$ & $ -8.82 $ & $ -9.26 $ & $ -2.41 $ & $ -2.89 $ & $ -6.65 $ \\
$\ln BF$ & $ -4.76 $ & $ -4.4 $ & $ 0.143 $ & $ -0.00586 $ & $ -0.0566 $ \\
\hline \hline
\end{tabular} }
\end{center}
\caption{ \textbf{$T^2$ cross section with $N_{\rm eff}$:} We report the 68\% (95\%) CL constraints/bounds on the cosmological parameters above the line, while below the line we have the improvement of the $\chi^2$ of the best fit and the Bayes Factor, with respect to the $\Lambda$CDM+$N_{\rm eff}$ scenario.  }
\label{tab4}
\end{table*}

In this appendix, we briefly discuss the results for the $T^2$-dependent $\nu$DM cross section. We report in \autoref{tab3} the constraints on the cosmological parameters at 68\% (and 95\% CL) for this case when $N_{\rm eff}$ is fixed to its standard value $3.044$.

Also in this scenario we have just an upper bound for the coupling between DM and neutrinos defined in \autoref{eq:defu}, in complete agreement with a model without interaction. In this case we have very similar bounds for Planck only and Planck +BAO, \textit{i.e.,} $\log_{10}u_{\nu \rm{DM}}<-15.4$ ($\log_{10}u_{\nu \rm{DM}}<-14.1$) at 68\% (95\%) CL, where $u_{\nu \rm{DM}} = [\sigma_{\nu\textrm{DM}}(T_0)/\sigma_{\textrm{T}}]\,(m_{\textrm{DM}}/100~\textrm{GeV})^{-1}$ corresponds to the present-day value of the scattering cross section. This interacting model results to be strongly favored by the value of the Bayes Factor reported in the last row of \autoref{tab3}.

Moreover, as it happens also in the temperature independent cross-section case, when we analyze the ACT data we have an indication at slightly more than 68\% CL for a coupling between DM and neutrinos different from zero and equal to $\log_{10}u_{\nu \rm{DM}}=-15.2^{+1.8}_{-1.1}$ at 1$\sigma$. 
However, also in this case, we find only an upper limit at 95\% CL. The results are  stable under the inclusion of the BAO measurements.

Interestingly, once we include in the analysis the Planck data up to multipoles $\ell=650$, and we study the combination ACT+Planck+BAO (last column of \autoref{tab3}) the 1$\sigma$ indication for a $\nu$DM coupling is slightly strengthened,
%and the peak more pronounced, as we can see in the right panel of \autoref{fig:1d}),
but this dataset combination is not powerful enough to bound the $\nu$DM coupling with a stronger statistical evidence. In particular we find $\log_{10}u_{\nu \rm{DM}}=-14.9^{+1.5}_{-0.63}$ at 68\% CL.

Also in this case, as a further step, we consider $N_{\rm eff}$ free to vary, and we report in \autoref{tab4} the constraints on the cosmological parameters at 68\% (and 95\% CL) for the $T^2$ $\nu$DM cross-section case. However, as already happened for the temperature independent cross section case, the constraints on $N_{\rm eff}$ obtained in this scenario are the same as those obtained assuming a non-interacting model, and all the features regarding the coupling in \autoref{tab3} remain valid also in this extended scenario.

Therefore, we find only an upper limit for $\log_{10}u_{\nu \rm{DM}}$ from Planck and Planck+BAO, and 1$\sigma$ indication for a DM-$\nu$ coupling different from zero for ACT, ACT + BAO, and ACT + BAO + the low-multipoles Planck data. In particular, we find for the full dataset combination ACT+Planck+BAO data $\log_{10}u_{\nu \rm{DM}}=-15.1^{+1.7}_{-0.90}$ at 68\% CL.

To conclude, when the full ACT+Planck+BAO dataset combination is considered, this interacting model fits the cosmological data as well as the standard $\Lambda$CDM scenario, and the two models are indistinguishable from a model comparison point of view, as we can deduce with the Bayes Factor in the last row of \autoref{tab3} and \autoref{tab4}. However, in general, the effects on the CMB anisotropies predicted by a the temperature dependent case are smaller, so that such the preference for a non-zero $\nu$DM interaction is somewhat mitigated with respect to the temperature-independent case (see also \hyperref[sec.Tplot]{Appendix B} where the correlation between cosmological parameters is shown for all the cases under study in this work). Therefore, we advise caution when interpreting the results for the temperature-dependent cross-section and suggest that this case warrants further analyses.

\section{Triangular Plot}
\label{sec.Tplot}

We show here the triangular plots for all the cases analysed involving the ACT data.

\begin{figure*}
    \centering
    \includegraphics[width=\textwidth]{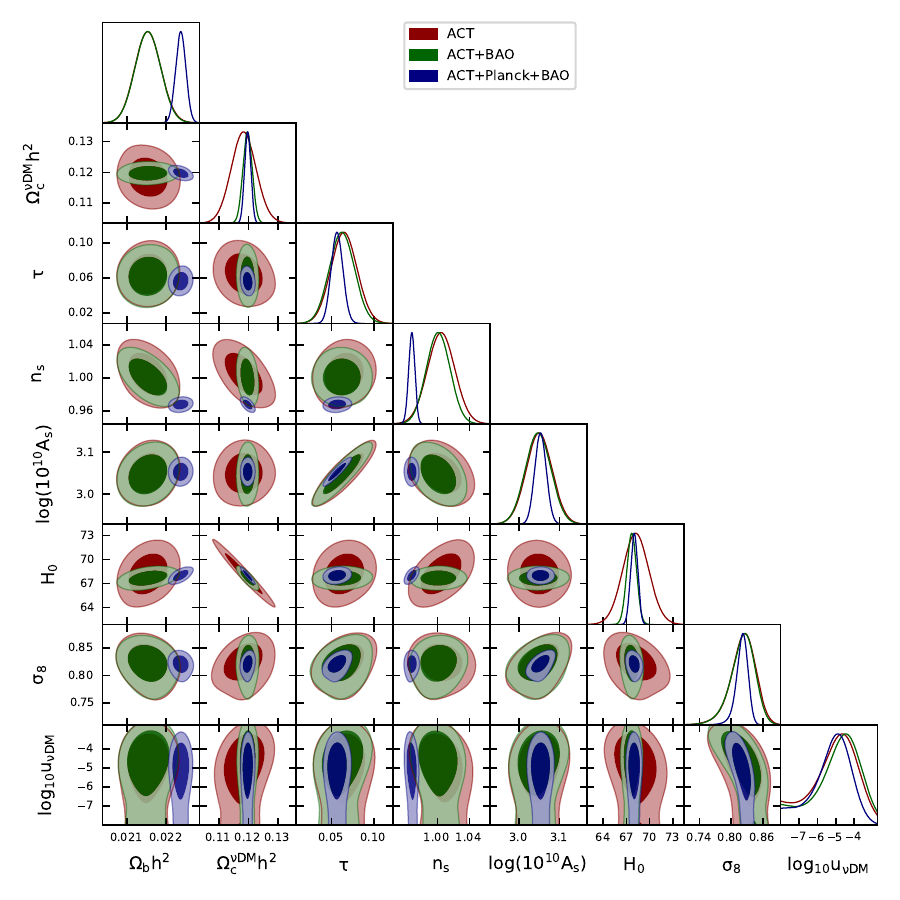}
    \caption{\textbf{Temperature independent cross section:} 1-dimensional marginalized posterior distributions and the 2-dimensional joint contours inferred for the most relevant parameters by analyzing the the Atacama Cosmology Telescope (ACT) CMB observations and their combination with BAO and Planck 2018}
    \label{fig:T0}
\end{figure*} 

\begin{figure*}
    \centering
    \includegraphics[width=\textwidth]{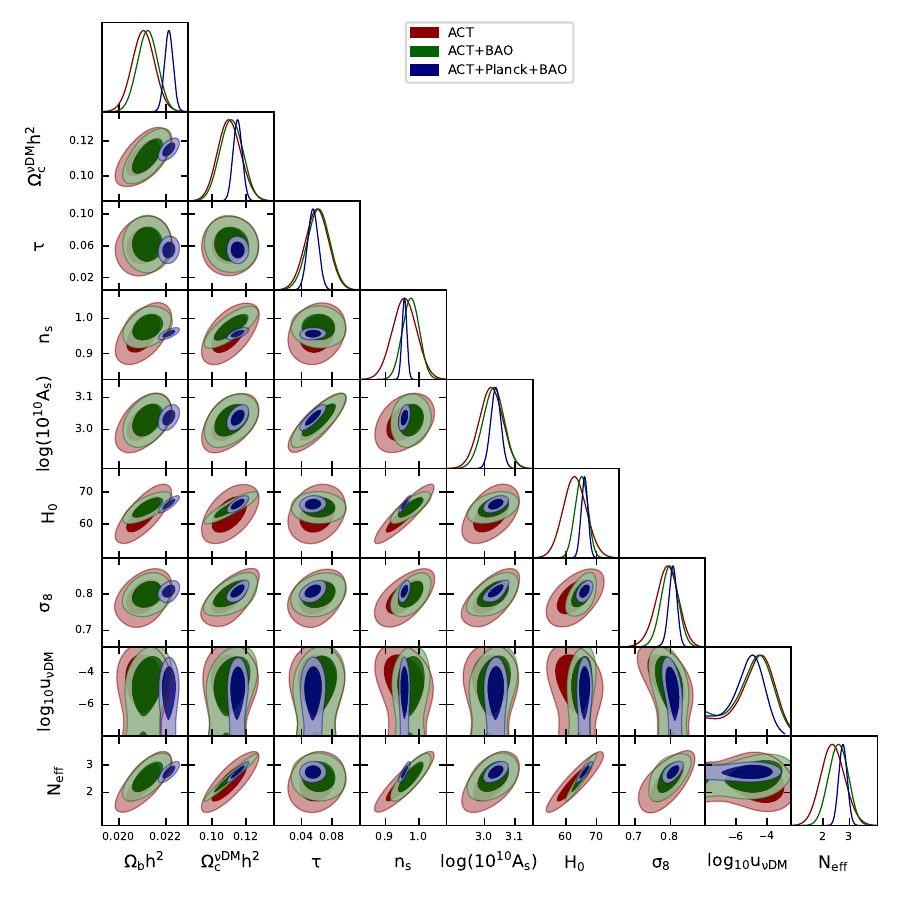}
    \caption{\textbf{Temperature independent cross section with $N_{\rm eff}$:} 1-dimensional marginalized posterior distributions and the 2-dimensional joint contours inferred for the most relevant parameters by analyzing the Atacama Cosmology Telescope (ACT) CMB observations and their combination with BAO}
    \label{fig:T0+Neff}
\end{figure*} 

\begin{figure*}
    \centering
    \includegraphics[width=\textwidth]{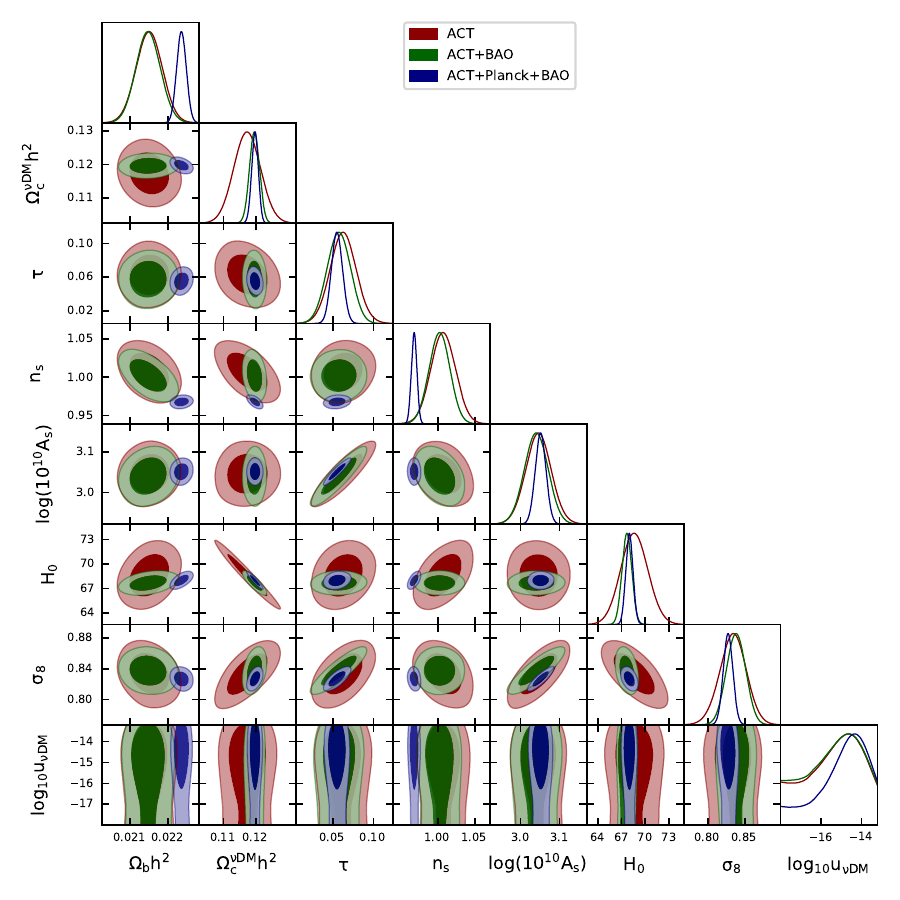}
    \caption{\textbf{$T^2$ cross section:} 1-dimensional marginalized posterior distributions and the 2-dimensional joint contours inferred for the most relevant parameters by analyzing the Atacama Cosmology Telescope (ACT) CMB observations and their combination with Planck and BAO}
    \label{fig:T2}
\end{figure*}

\begin{figure*}
    \centering
    \includegraphics[width=\textwidth]{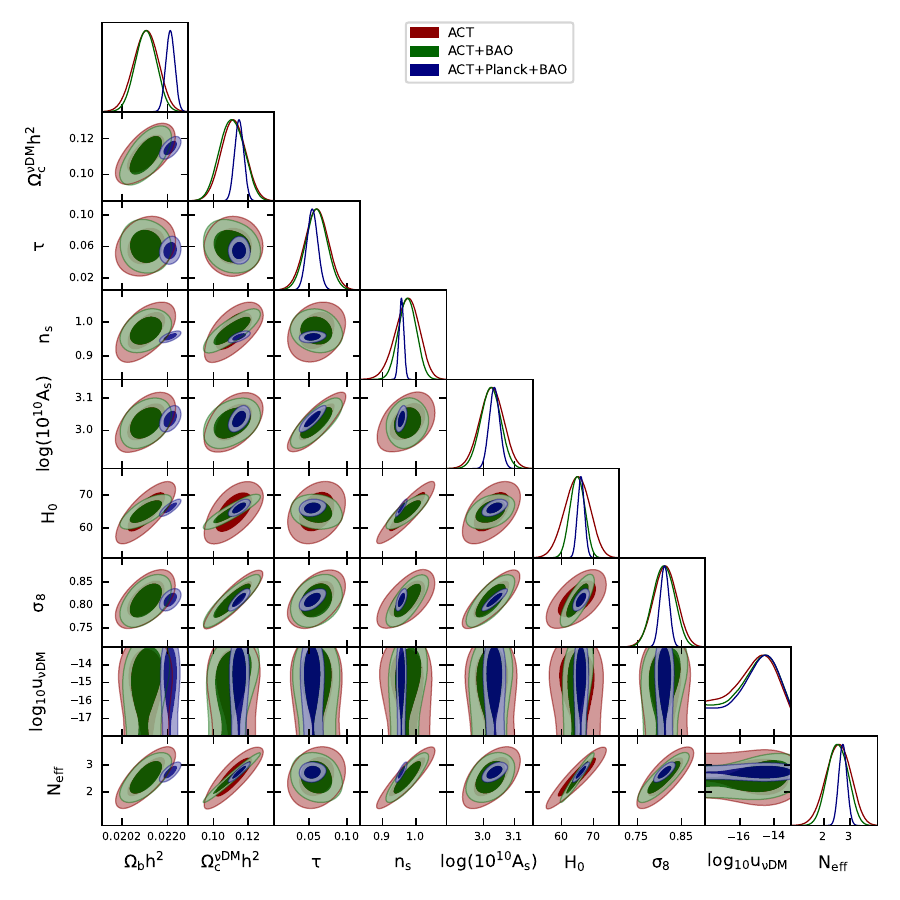}
    \caption{\textbf{$T^2$ cross section with $N_{\rm eff}$:} 1-dimensional marginalized posterior distributions and the 2-dimensional joint contours inferred for the most relevant parameters by analyzing the Atacama Cosmology Telescope (ACT) CMB observations and their combination with Planck and BAO}
    \label{fig:T2+Neff}
\end{figure*}

\clearpage
\bibliography{main}

\end{document}